\documentclass[preprint,showpacs,preprintnumbers,amsmath,amssymb,superscriptaddress]{revtex4}

\usepackage{graphicx,graphics}   

\usepackage[utf8]{inputenc}
\usepackage{graphicx,amsfonts}
\usepackage{epstopdf}
\usepackage{subfig}
\usepackage{cancel}
\usepackage{float}
\usepackage[english]{babel}
\usepackage{dcolumn}
\usepackage{bm}
\begin{document}

\title{ Neutral meson mixing induced by box diagrams in the 3-3-1 model with heavy leptons}

\author{F. C. Correia}%
\email{ccorreia@ift.unesp.br}
\affiliation{
Instituto  de F\'\i sica Te\'orica--Universidade Estadual Paulista \\
R. Dr. Bento Teobaldo Ferraz 271, Barra Funda\\ S\~ao Paulo - SP, 01140-070,
Brazil
}
\author{V. Pleitez}%
\email{vicente@ift.unesp.br}
\affiliation{
Instituto  de F\'\i sica Te\'orica--Universidade Estadual Paulista \\
R. Dr. Bento Teobaldo Ferraz 271, Barra Funda\\ S\~ao Paulo - SP, 01140-070,
Brazil
}

\date{08/27/15}
%
\begin{abstract}
We consider in the 3-3-1 model with heavy leptons the box contributions to the mass difference  in $K$ and $B$ neutral mesons induced by neutral (pseudo)scalars, exotic charged quarks, singly and doubly charged scalar and gauge bosons. In particular, we include the effects of a real scalar with mass near 125 GeV but with non-diagonal couplings to quarks. We show that, as in the tree level case, there are ranges of the parameters in which these contributions can be enough suppressed by negative interference among several amplitudes. Hence, in this model these $\Delta F=2$ processes may be dominated by the standard model contributions. In addition, our results are valid in the minimal 3-3-1 model without the sextet.  
\end{abstract}

\pacs{12.15.Mm 
12.60.Cn 
12.15.Ji 
}

\maketitle

\section{Introduction}
\label{sec:o}

Nowadays all the prediction of the Standard Model (SM) have been experimentally tested and are in agreement with the model's predictions at a given order in perturbation theory. However, there are reasons for expecting the existence of new particles. Among others, the existence of dark matter~\cite{Ade:2013zuv} and the neutrino masses~\cite{Agashe:2014kda} need, for their implementation in any model, particles that do not belong to the degrees of freedom in the SM. These new particles, if any, can be observed by the direct search at colliders like the LHC, or by their effects on rare decays that are suppressed, by several reasons, in the SM~\cite{Buras:2015nta}. For instance, the rare $B^0_s\to\mu^+\mu^-$ decay has been recently observed at CERN~\cite{CMS:2014xfa} with a branching ratio compatible with the SM model prediction. 

Almost all the extensions of the electroweak standard model (ESM) have a rich scalar sector. Some of them introduce new quarks and/or leptons along with neutral and charged extra scalars and vector bosons. 
All these cases occur in models whose gauge symmetry is larger than the SM symmetries, in particular in the minimal 3-3-1 model (m331 as shorthand)~\cite{Pisano:1991ee,Frampton:1992wt,Foot:1992rh}, and in the 3-3-1 model with heavy leptons (331HL)~\cite{Pleitez:1992xh}. 
Moreover, the extra neutral vector (generically denoted by $Z^\prime$) and scalar bosons induce flavor changing neutral~currents (FCNCs) processes at tree level which are supposed to be the dominant extra contributions (besides that of the SM). In 331 models, as in many extensions of the SM, such processes are induced by neutral (pseudo) scalars and, since the discovery of a spin-0 resonance with mass 125 GeV~\cite{Aad:2012tfa,Chatrchyan:2012ufa}, it is mandatory to take its effects into account.

The FCNC at the tree level were revisited in the context of the m331 model~\cite{Machado:2013jca} for two reasons: Firstly, usually when considering the $Z^\prime$ phenomenology in this model, the contributions of the (pseudo)scalars are neglected~\cite{Rodriguez:2004mw,CarcamoHernandez:2005ka,Promberger:2007py,Godfrey:2008vf, Duenas:1900zz,Barreto:2010ir,Cabarcas:2009vb,Cabarcas:2011hb,Buras:2012dp}. Secondly, as we said before, the scalar sector of any model beyond the SM (BSM) must contain a scalar field with the mass around 125 GeV and diagonal couplings compatible, within the experimental error, with those of the ESM Higgs boson at least with the third quark generation. However, generally this ESM-like Higgs scalar also mediate FCNC at tree level and its effects have to be computed.  See for instance~\cite{Blankenburg:2012ex}. From the experimental point of view, the CMS has reported measured of  $h\to \mu\tau$ which is 2.5$\sigma$ different from zero~\cite{Khachatryan:2015kon}. Recall that in the ESM, FCNC processes occur only at the 1-loop level~\cite{Benitez-Guzman:2015ana}. 

The conclusions of Ref.~\cite{Machado:2013jca} establish that in $\Delta F=1,2$ FCNC processes, when the $C\!P$ even SM-like neutral Higgs boson and one of the $C\!P$ odd scalars are considered, there are positive and negative interference among these fields and the $Z^\prime$ in such a way that the previous constraints on the mass of the $Z^\prime$ boson are avoided.  For instance,  the measured value of strange and bottom mesons, $\Delta M_{K,B_{(s)}}$, a lower limit  $M_{Z^\prime}>1.8$~TeV is still possible and is also compatible with the constraints coming from weak decays.  Depending on the values of the unitary matrices in the neutral scalar and pseudo-scalar sectors, even lower values may be allowed. 

The value of the lower limit for the $Z^\prime$ mass depends not only on the projections of the neutral scalars over the SM-like Higgs but also on the unitary matrices which rotate quarks and leptons symmetry eigenstates to the respective mass eigenstates, $V^{U,D}_{L,R}$. Nu\-me\-ri\-cal values for the latter matrices were obtained in~\cite{Machado:2013jca}. This reduces the number of free parameters in the model. In fact, the only free parameters remaining are the unitary matrices that diagonalize the scalar mass matrices. These results are valid in the m331 and in 331HL models since both have the same quark content. However the scalar mass spectra are different in both models, the former needs a scalar sextet for generating the charged leptons masses, and the later one does not. 

Here we will concentrate only in the 331HL model. Our results are valid in the m331 when the sextet is avoided and the charged lepton masses need the contributions of a dimension five operator built with two triplets~\cite{DeConto:2015eia}. We emphasize that the entries of the matrices $V^{U,D}_{L,R}$ obtained in~\cite{Machado:2013jca} are not unique and different solutions imply different phenomenology. 

In Ref.~\cite{Machado:2013jca} it was reasonable to consider only the tree level amplitudes because certainly these are the main extra contributions to the $\Delta F=1,2$ processes in the m331 model and in addition,  these amplitudes involve only the $Z^\prime$, neutral scalars and pseudo-scalars. However, there are also 1-loop diagrams, for instance boxes and penguin, which include not only the neutral scalar and vector bosons, but all the particle spectrum of the model: exotic charged quarks, singly and double charged scalar and vector bosons. Accordingly, we have to evaluate their effects in order to see if there is, or not, some values of the parameters of the model in which the constraints on the mass of $Z^\prime$ obtained at tree level are not spoiled by 1-loop corrections. Hence, it is necessary to quantify the effects of all these particles in FCNC processes and this is the aim of the present paper. 

The outline of this paper is as follows. In Sec.~\ref{sec:model} we present the representation content of the model. In Subsec.~\ref{subsec:yukawa} and \ref{subsec:cc} we show the Yukawa and the quark-vector boson interactions, respectively. In Sec.~\ref{sec:eh} we give the effective Hamiltonian which arises from the boxes with at least one of the extra particles in the 331HL model. In Subsec.~\ref{subsec:qs} we consider the amplitudes involving two (pseudo)scalars, while boxes with two vector bosons are considered in Subsec.~\ref{subsec:qvb}. Those with one scalar and one vector bosons are shown in Subsec.~\ref{subsec:qvs}. Boxes involving one photon or one $Z$ is shown in Subsecs:~\ref{subsec:photon} and \ref{subsec:Z}, respectively. Our results are summarized in Sec.~\ref{sec:mesons} while the last section is devoted to our conclusions. In the Appendixes we consider: in Appendix~\ref{sec:matrices} we write explicitly the matrices appearing in Subsecs.~\ref{subsec:qs} -- \ref{subsec:qvb}. In Appendix~\ref{sec:integrals} we show the types of integrals which arise from the boxes considered. The scalar mass spectra and mass eigenstates are summarized in Appendix~\ref{sec:scalarshl}; while the matrix elements in the vacuum insertion approximation are given in Appendix~\ref{sec:vi}. Finally, in the Appendix~\ref{sec:examples} we give four examples of the amplitudes that have been calculated in this work: the case of the exchange of $Y^+_1$ and a Goldstone boson $G^+_V$ in Appendix~\ref{subsec:21}, the exchange of a $Y^-_1$ and a $W^-$ in Appendix~\ref{subsec:22}, and that of the exchange of two charged vector bosons is shown in Appendix~\ref{subsec:23}. In Subsec.~\ref{subsec:24} we shown that the penguin-like diagrams are negligible, at least with the values of $V^{U,D}_{L,R}$ used in this paper. 

\section{The model}
\label{sec:model}

In the 331HL model~\cite{Pleitez:1992xh}  the
left-handed quark fields are chosen to form two anti-triplets
$Q^\prime_{mL}=(d_{m}\, -\!\!u_{m}\;J_{m})_{L}^{T}\sim({\bf
3}^{*},-1/3);\; m=1,2$; and a triplet $Q^\prime_{3L}=(u_{3}\, d_{3}\,
J_3)_{L}^{T} \sim({\bf 3},2/3)$ and the right-handed ones are in singlets: $u_{\alpha
R}\sim({\bf 1},2/3)$, $d_{\alpha R}\sim({\bf
1},-1/3),\,\alpha=1,2,3$, $J_{mR}\sim({\bf 1},-4/3)$, and
$J_{3R}\sim({\bf 1},5/3)$.  Below we will use $J_3\equiv J$.
The numbers between parenthesis mean the transformation properties under $SU(3)_{L}$
and $U(1)_{X}$, respectively. 
We have omitted $SU(3)_C$ factor because all quarks are triplets of $SU(3)_C$. Lepton generations are all in triplets $\Psi_L=(\nu_\ell\, \ell \, E_\ell)^T_L\sim(\textbf{3},0)$ and right handed charged lepton fields $\ell_R\sim(\textbf{1},-1)$ and $E_{\ell R}\sim(\textbf{1},+1)$. In the scalar sector we have three triplets: $\eta=(\eta^0\,\eta^{-}_1\,\eta^+_2)^T\sim({\bf3},0)$,
$\rho=(\rho^+\,\rho^0\,\rho^{++})^T\sim({\bf3},1)$,
$\chi=(\chi^-\,\chi^{--}\,\chi^0)^T\sim({\bf3},-1)$. 
Only the three scalar  triplets are needed to break the gauge symmetries and generate all the fermion  masses.
The model has, besides the photon, $W^\pm_\mu$ and $Z_\mu$, an extra neutral vector boson, $Z^\prime_\mu$,
and single and doubly charged bileptons, generically denoted by $V^\pm_\mu$ and $U^{\pm\pm}_\mu$ when they are vectors, and $Y^\pm_{1,2}$ and $Y^{\pm\pm}$, when they are scalars. 

\subsection{Yukawa interactions}
\label{subsec:yukawa}

The Yukawa interactions in the quark sector are given by:
\begin{eqnarray} 
-\mathcal{L}^q_Y &=& \bar{Q}_{mL} \left[ G_{m\alpha} U^\prime_{\alpha R} \rho^*+\tilde{G}_{m\alpha}D^{'}_{\alpha R} \eta^* \right]+
\bar{Q}_{3L} \left[ F_{3\alpha}U^\prime_{\alpha R} \eta + \tilde{F}_{3\alpha}D^{'}_{\alpha R} \rho \right] \nonumber \\&  +&
\bar{Q}_{mL}G^\prime_{mn}J_{nR}\chi^* + \bar{Q}_{3L}g_J J_R \chi +H.c.,
\label{q1}
\label{quarks}
\end{eqnarray}
where we omitted the sum in $m,n=1,2$, and $\alpha=1,2,3$, $U^\prime_{\alpha R}=(u^\prime\,c^\prime\,t^\prime)_R$ and $D^{'}_{\alpha R}~=~
(d^\prime\,s^\prime\,b^\prime)_R$. $G_{m\alpha}$, $\tilde{G}_{m\alpha}$, $F_{3\alpha}$, $\tilde{F}_{3\alpha}$,  are the coupling constants whose values were obtained in Ref.~\cite{Machado:2013jca} and are reproduced in Appendix~\ref{sec:matrices}. The $2\times2 $ matrix $G^\prime_{mn}$ and $g_J$ are all free Yukawa couplings which determine the mass of the exotic quarks. The mass matrix of the quarks with electric charge $-4/3$, $J_1,J_2$, is diagonalized by an orthogonal matrix and 
\begin{equation}
\left(\begin{array}{c}
J_1\\J_2
\end{array} 
\right) =\left(
\begin{array}{cc}
\cos\theta & \sin\theta \\
-\sin \theta & \cos\theta 
\end{array}\right)
\left(\begin{array}{c}
j_1\\j_2
\end{array} 
\right),
\label{jotas}
\end{equation}
where $\theta$ is a new mixing angle in the model and $j_{1,2}$ are mass eigenstates.  

From Eq.~(\ref{quarks}), we obtain the interactions involving quarks and charged scalars are
\begin{eqnarray}
- \mathcal{L} _{\textbf{S}} &=& \overline{j}_L [
\tilde{K}_1\,  G_V ^- 
+ \tilde{K}_2\,   Y_2 ^-] D_R
+ \overline{D}_L [ K_2\, Y_2 ^+ 
+ K_1\,   G_V ^+ ]j_R
\nonumber \\
& & + \overline{\mathcal{J}}_L [
\tilde{K}_3\, G_U^{++}
+ \tilde{K}_4\,   Y ^{++}]D_R + 
\overline{D}_L  [ K_4    Y^{--}
+ K_3\,  G_U^{--}] \mathcal{J}_R  + \nonumber \\
& & + \overline{D}_L  [K_5\,   G_W^- + K_6\,  Y_1 ^-]  U_R 
+ \overline{U}_L  [\tilde{K}_6 \,  Y_1 ^+ + \tilde{K}_5\,   G_W^+] D_R + h.c.  
\label{esc1}
\end{eqnarray}
where $j_L = (J_1 \ J_2 \ 0)_L$ and $\mathcal{J}_{L,R} = (0 \ 0 \ J)_{L,R}$,  $U^T_{L,R}=(u\,c\,t)_{L,R}$ and $D^T_{L,R}=(d\,s\,b)_{L,R}$ are quarks mass eigenstates. We write the vertices in terms of the symmetry eigenstaes $J_{1,2}$ but the decomposition in Eq.~(\ref{jotas}) has been considered in the calculations.
Our calculations are performed in the Feynman-'t Hooft gauge. Thus, in Eq.~(\ref{esc1}), $G_{V,U,W}$ denote the would-be Goldstone bosons related to $V^\pm,U^{\pm\pm}$ and $W^\pm$, respectively.

\begin{center}
	\begin{figure}[!ht]
		\subfloat{\includegraphics[width=7cm]{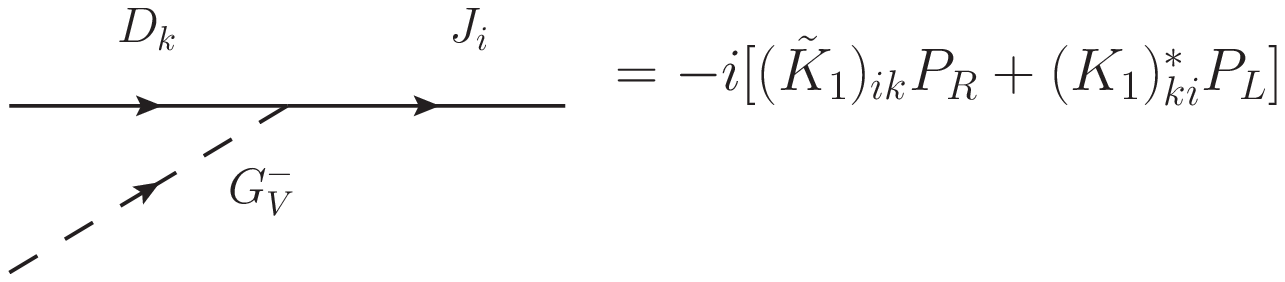}} \ \ \
		\subfloat{\includegraphics[width=7cm]{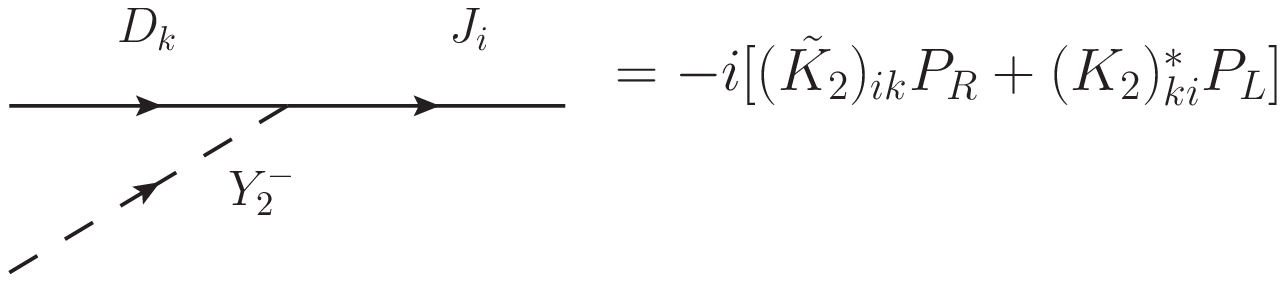}} \\
		\subfloat{\includegraphics[width=7cm]{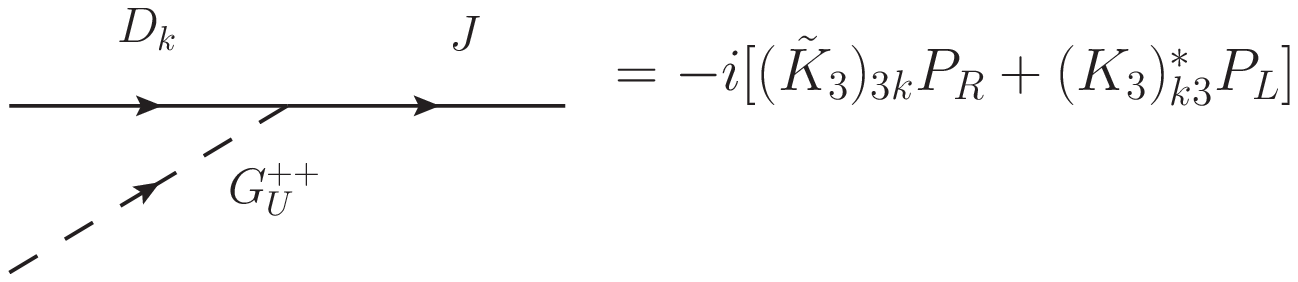}} \ \ \
		\subfloat{\includegraphics[width=7cm]{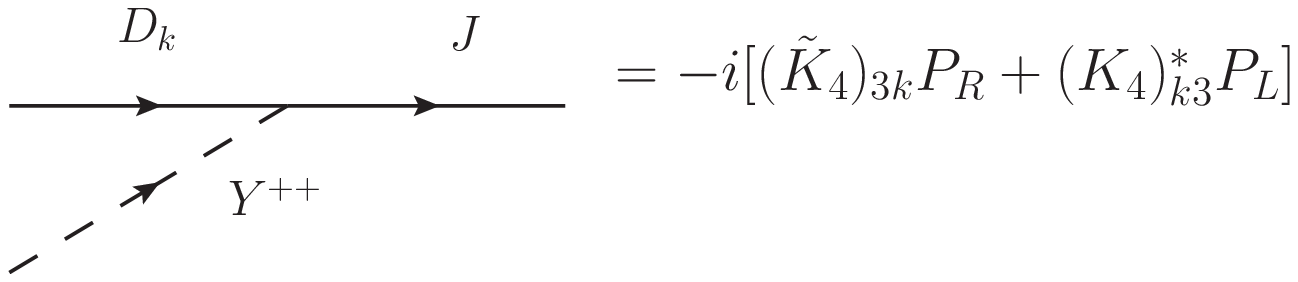}} \\
		\subfloat{\includegraphics[width=7cm]{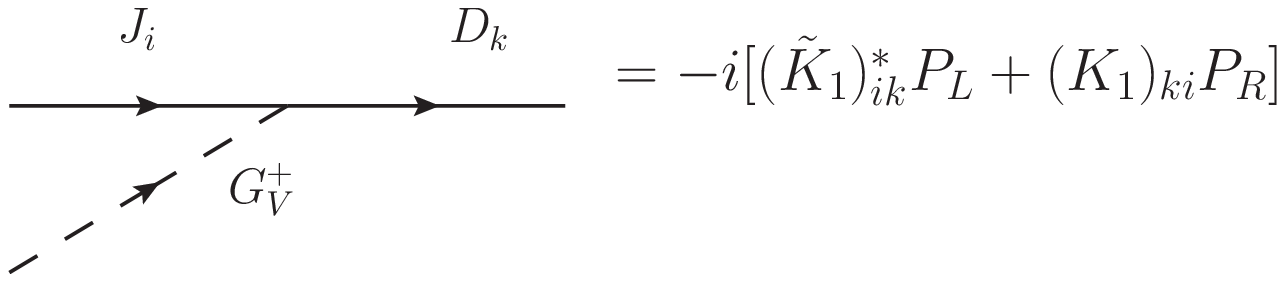}} \ \ \
		\subfloat{\includegraphics[width=7cm]{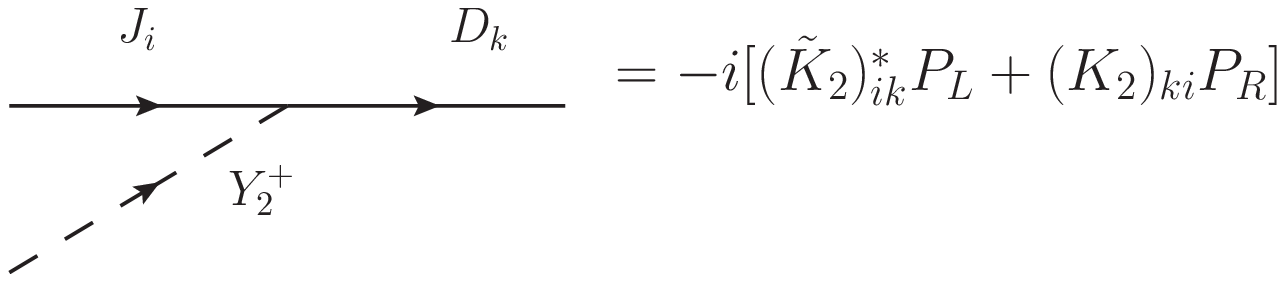}} \\
		\subfloat{\includegraphics[width=7cm]{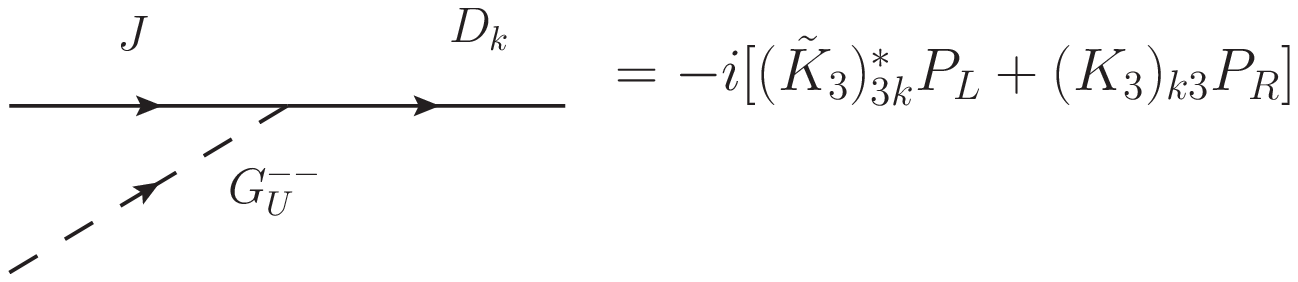}} \ \ \
		\subfloat{\includegraphics[width=7cm]{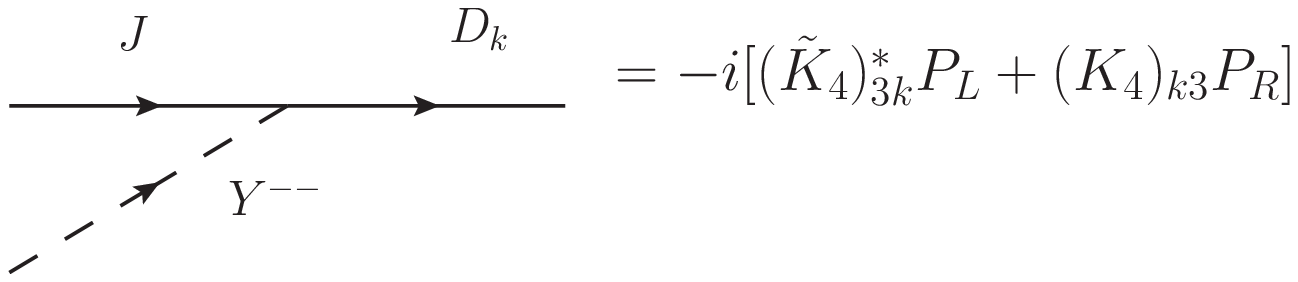}} \\
		\subfloat{\includegraphics[width=7cm]{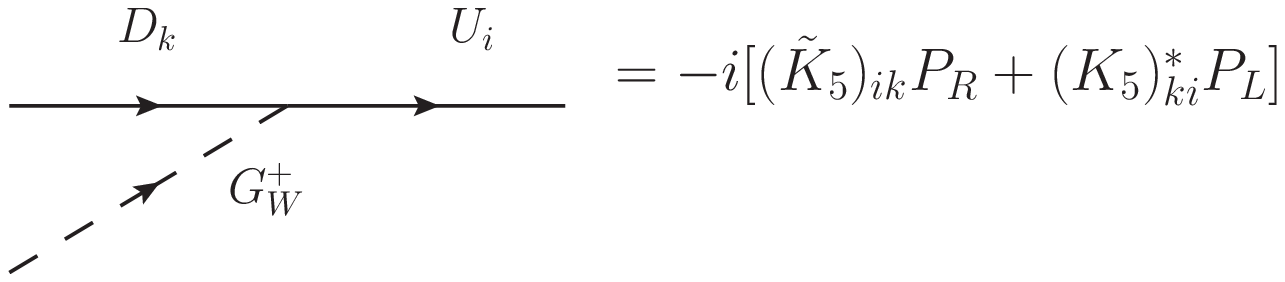}} \ \ \
		\subfloat{\includegraphics[width=7cm]{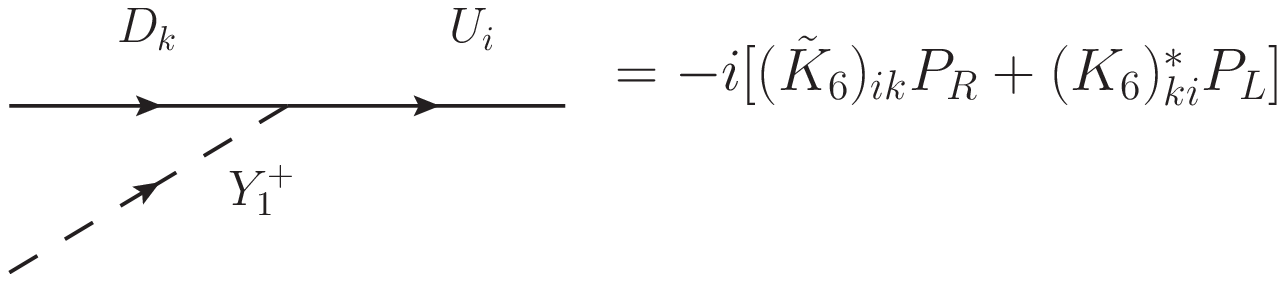}} \\
		\subfloat{\includegraphics[width=7cm]{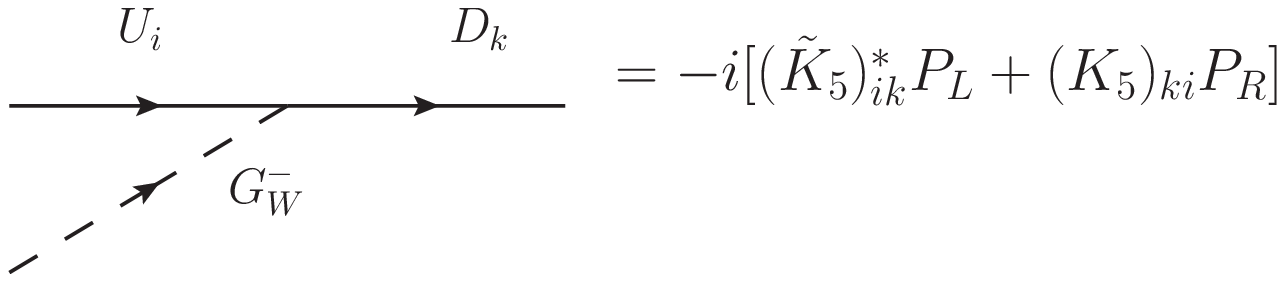}} \ \ \
		\subfloat{\includegraphics[width=7cm]{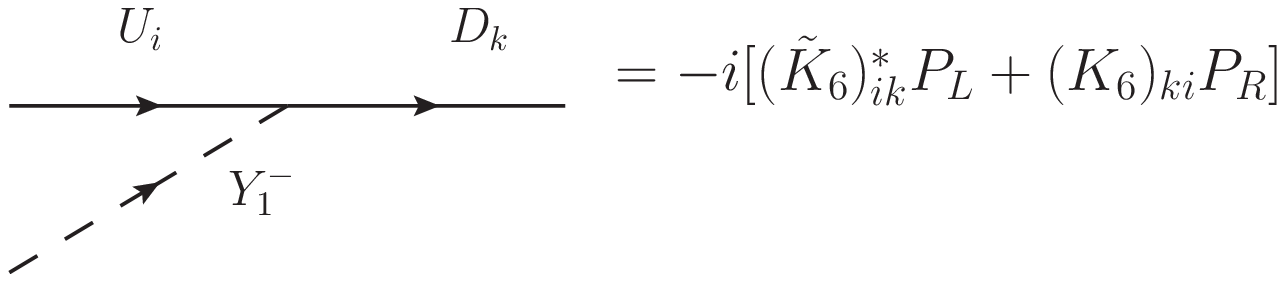}}  
		\caption{Vertices of quarks and charged scalars. Here $J_i,\,i=1,2$ denotes the symmetry eigenstates in Eq.~(\ref{jotas}).}\label{fig1}
	\end{figure}
\end{center}

\begin{center}
	\begin{figure}[H]
		\subfloat[]{\includegraphics[width=7cm]{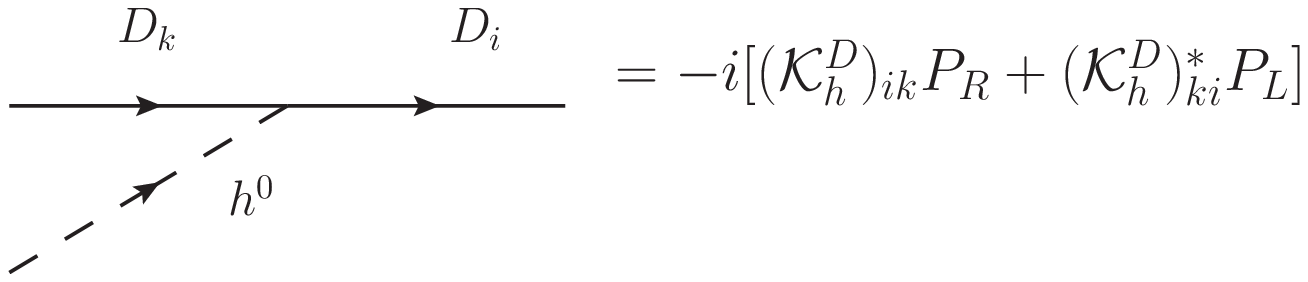}} \ \ \
		\subfloat[]{\includegraphics[width=7cm]{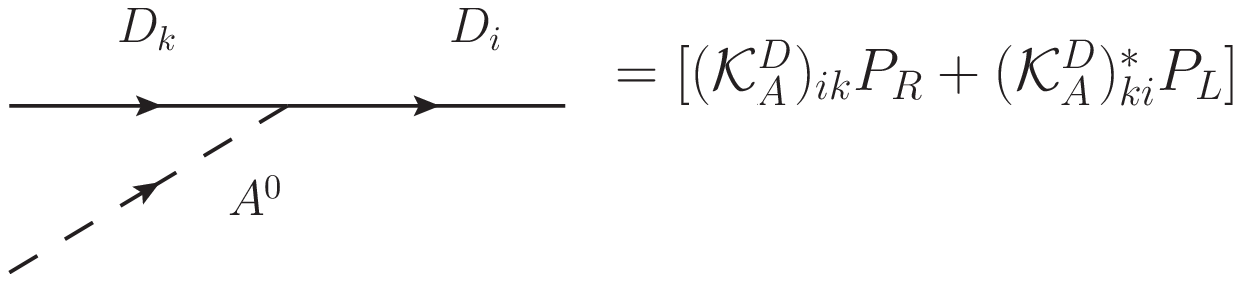}}
		\caption{Vertices of quarks and neutral scalars.}\label{fig2}
	\end{figure}
\end{center}

\begin{center}
 	\begin{figure}[H]
 		\subfloat[]{\includegraphics[width=7cm]{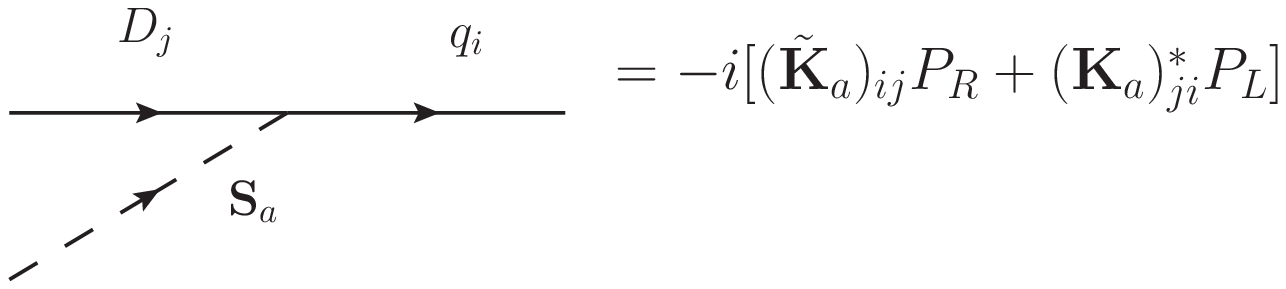}} \ \ \ 
 		\subfloat[]{\includegraphics[width=7cm]{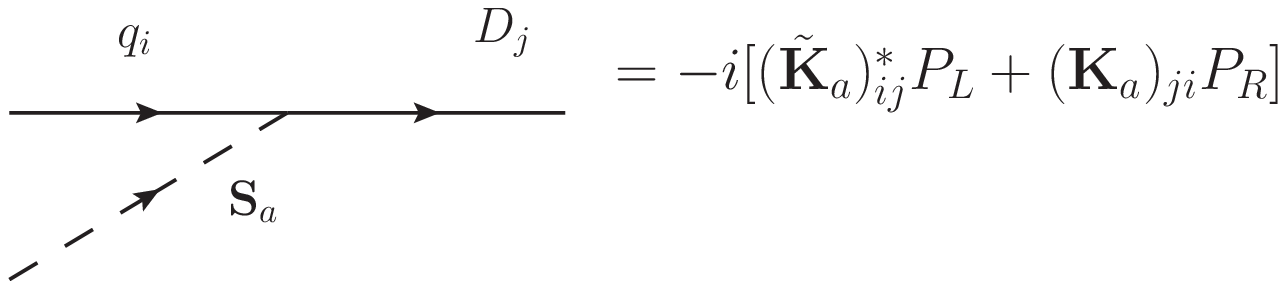}}
 		\caption{The generic vertices to quarks and scalars.}\label{fig3}
 	\end{figure}
 \end{center}

Analogously, the interaction of quarks and neutral scalar can be written as
\begin{equation}
-\mathcal{L}_{\textbf{S}^0}= \overline{D}_L \mathcal{K}^D_h D_R h^0  + 
i \overline{D}_L \mathcal{K}^D_A D_R A^0 + H.c.
\label{esc2}
\end{equation}
The matrices $K_a, \tilde{K}_a,\,a=1,\cdots,6$, appearing in (\ref{esc1}) are given in Eqs.~(\ref{a1e2})  - (\ref{a1e4}) in the Appendix~\ref{sec:matrices}, while $\mathcal{K}^D_h$ and $\mathcal{K}^D_A$ in (\ref{esc2}) are shown in Eq.~(\ref{a1e5}).
The respective vertices are obtained as usual by separating the constants in $i\mathcal{L}$ and represented in Figs.~\ref{fig1} and Fig.~\ref{fig2}. The generic vertex is written as in Fig.~\ref{fig3}.

In Eq.~(\ref{esc2}) we have already assumed that the $\rho^0$ is the neutral scalar which has the largest projection on the scalar whose mass is around $125$ GeV. In Ref.~\cite{Machado:2013jca} was shown that, if $U_{\rho1}=0.42$, this scalar has the same couplings with the top and b-quark as in the SM.

\subsection{Fermion-vector boson interactions}
\label{subsec:cc}

\begin{center}
 	\begin{figure}[!ht]
 		\subfloat{\includegraphics[width=7cm]{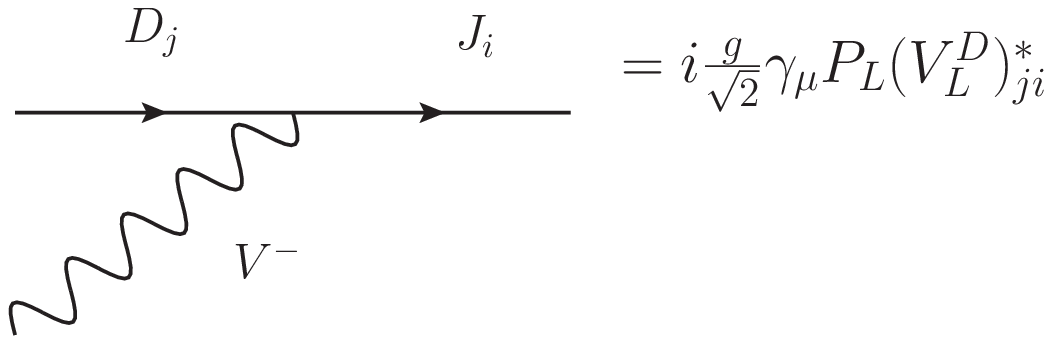}} \ \ \
 		\subfloat{\includegraphics[width=7cm]{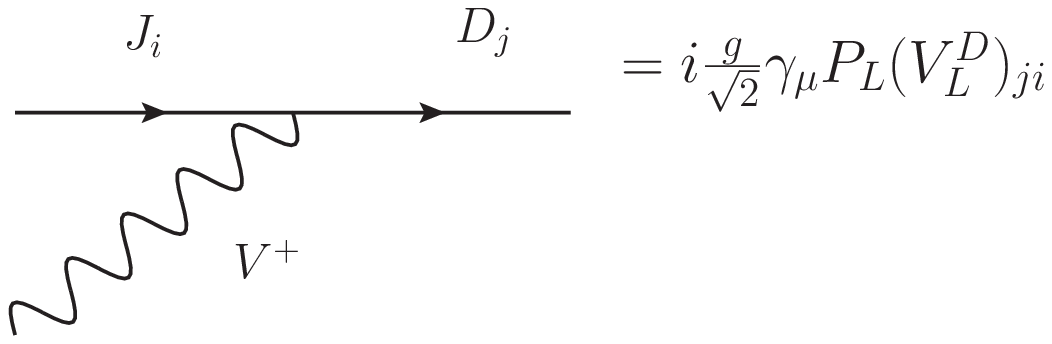}} \ \ \
 		\subfloat{\includegraphics[width=7cm]{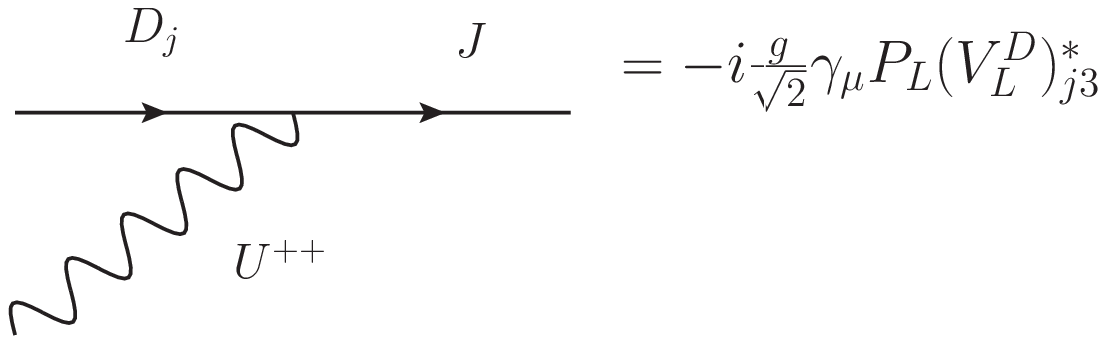}} \ \ \
 		\subfloat{\includegraphics[width=7cm]{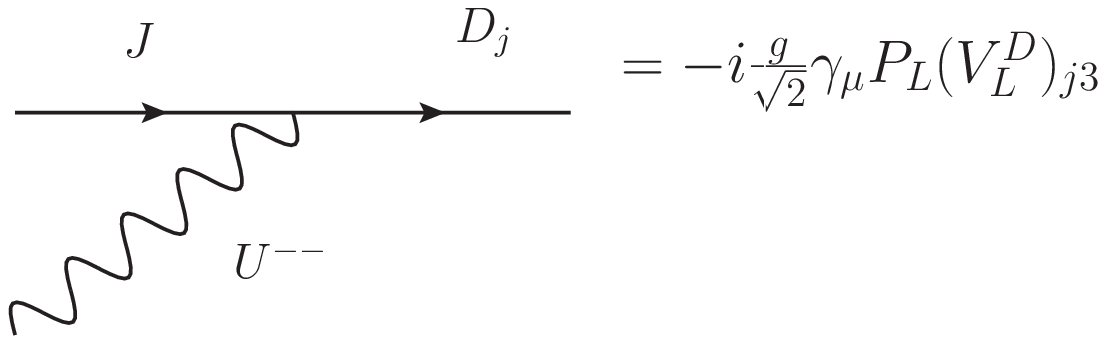}} 
 		\caption{Vertices of quarks and gauge bosons. Here $J_i,\,i=1,2$ denotes the symmetry eigenstates in Eq.~(\ref{jotas}).}\label{fig4}
 	\end{figure}
 \end{center}

\begin{figure}[ht]
  \includegraphics[scale=0.6]{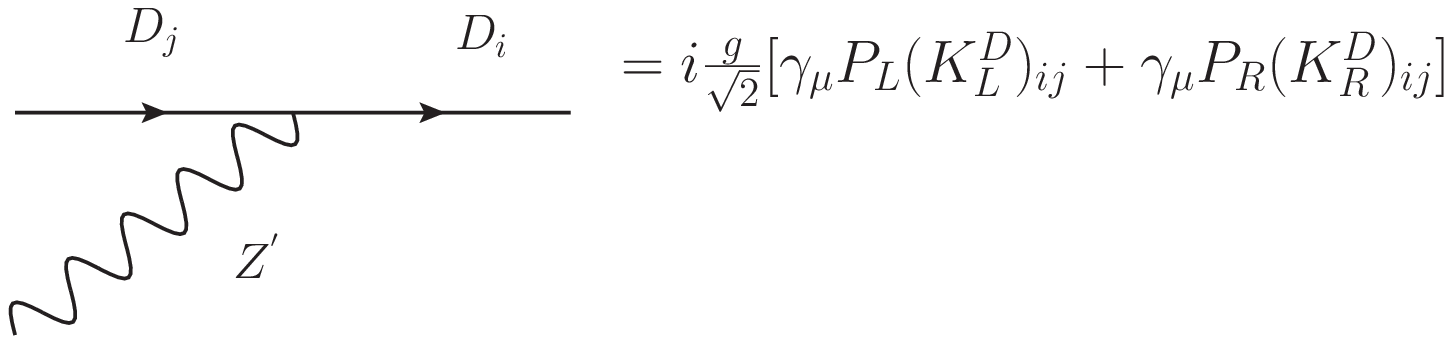}
  \caption{Interaction vertex of quarks and $Z'$.} \label{fig5}
  \end{figure}

The interactions between vector bosons $V^-_\mu,U^{--}_\mu, Z^\prime_\mu$ with quarks are:
\begin{equation}
\mathcal{L}_{\textbf{VB}} = \frac{g}{\sqrt{2}}[\mathcal{L}_V+\mathcal{L}_U+\mathcal{L}_{Z^\prime}],
\label{lvb}
\end{equation}
where 
\begin{eqnarray}
&&\mathcal{L}_V=  \bar{D_i} \gamma^{\mu} P_L 
(V_L ^D)_{im} J_m  V^+_\mu + \bar{J}_m \gamma^\mu P_L (V_L ^D)^* _{im} D_i  V^-_\mu, 
\nonumber \\&& 
\mathcal{L}_U= - \left[\bar{D_i} \gamma^\mu P_L (V_L ^D)_{i3} J  U^{--}_\mu + \bar{J} \gamma^\mu P_L (V_L ^D)^* _{i3} D_i  U^{++}_ \mu \right],
\nonumber \\&& 
\mathcal{L}_{Z^\prime}=\left[ \bar{D_i} \gamma^\mu P_L 
(K^D_L)_{ij} D_j + \bar{D_i} \gamma^\mu P_R 
(K^D_R)_{ij} D_j \right] Z^\prime_\mu.
\label{zprime}
\end{eqnarray}
and in $\mathcal{L}_V$ we have to sum over $i=1,2,3;m=1,2$, in $\mathcal{L}_U$ we sum over $i=1,2,3$ 
and in $\mathcal{L}_{Z^\prime}$ we sum over $i,j=1,2,3$. As before, $U^T_{L,R}=(u\,c\,t)_{L,R}$, $D^T_{L,R}=(d\,s\,b)_{L,R}$  and $P_L=(1-\gamma_5)/2$ and $P_R=(1+\gamma_5)/2$ are the chiral projectors. Here $(K^D_R)_{ij}~=~(2s^2_W/\sqrt{3}\sqrt{1-4s^2_W})\,\delta_{ij}\approx~0.97232\,\delta_{ij}$, and the matrix $K^D_L$ is shown in Eq.~(\ref{a1e6}). At this stage all quarks are already mass eigenstates (unprimed fields). We note that the interactions of the right-handed quarks with $Z^\prime$ conserve the flavor, however there are diagrams with two $Z^\prime$ in which one of them has right-handed vertices and the second left-handed ones.  We have omitted the neutral interactions of the $u$-type quarks but they involve similar matrices $K^U_L$ and $K^U_R$. The latter is also proportional to the unit matrix. We note that $K^{D,U}_L$ are not unitary matrices, for their definition see Ref.~\cite{Machado:2013jca}.
The respective vertices are given in Fig.~\ref{fig4} and \ref{fig5}. The generic vertex is shown in Fig.~\ref{fig6}.

\section{Effective Hamiltonian}
\label{sec:eh}

The new particle content in the 331HL model imply hundred of box diagrams and, at first, our main task will be to sort them in an irreducible way. 

The vertices derived from Eqs.~(\ref{esc1}) and (\ref{esc2}) are written in terms of the mass eigenstates and some matrices. We see from Figs.~\ref{fig1} and \ref{fig2} that there is a pattern which prevails about all the interactions and can be summarize as in Fig.~\ref{fig3}. The ${\textbf{K}}_a,\tilde{\textbf{K}}_a $  matrices will denote the matrices given in Eqs.~(\ref{a1e2})-(\ref{a1e5}), and are linked with the respective charged or neutral scalar, generically denoted by $\textbf{S}_a$. 

The next step is to find a general expression also for quark-gauge bosons interactions. As before, from Eq.~(\ref{lvb}), we extract the Figs.~\ref{fig4} and \ref{fig5}, and the results are shown in Fig.~\ref{fig6}. The matrix $\mathbf{V}$ denotes the matrices $V_L^D$ or $K^D_L$, in Eqs.~(\ref{a1e6}) and (\ref{a1e7}), whenever $\mathbf{VB}$ is a charged vector boson or the $Z^\prime$, respectively. The remaining terms due to $K^D_R$ will be explicitly written in  Eq.~(\ref{ampc}).

Hence, there are only two different generic vertices and their respective conjugates for all new interactions of 331HL. The laborious task of setting up a lot of new diagrams now reduces to figure out five independent diagrams, namely:
\begin{description}
\item[i)] those with two scalars denoted by $\textbf{S}_a\textbf{S}_b$, where $\textbf{S}_{a,b}$ run over all the (pseudo)scalars, neutral or charged. See Fig.~\ref{fig7}(a);
\item[ii)] diagrams with one vector boson $\textbf{VB}= V^\pm_\mu, U^{\pm\pm}_\mu, Z'$ and one scalar $\textbf{S}_a$, denoted by $\textbf{VBS}$. See Fig.~\ref{fig7}(b));
\item[iii)] with two vector bosons. See Fig.~\ref{fig8};
\item[iv)] with one photon, $\gamma$, and one neutral (pseudo)scalar or $Z'$. See Fig.~\ref{fig9}(b). 
\end{description}
We must remark that once the scalar or gauge boson are defined, the quark in the internal lines are also fixed. For example, if $\textbf{S}_a = \textbf{S}_b = Y^+_2$ the quarks will be $q_{i,l}=J_1, J_2$, if $\textbf{S}_a = \textbf{S}_b = h^0$ then $q_{i,l}=D_{i,l}$, and so on. 

We are interested in the effective interactions contributing to the mass difference of the pseudoscalar mesons 
$K$ and $B_{d,s}$. Therefore, after obtaining the amplitudes from our four kinds of diagrams we must sum over all the matrices:

\begin{equation}
\mathcal{H}^{\Delta F = 2}_{\textbf{eff}}=-\Bigl(\sum_{\textbf{S}_a\textbf{S}_b}\mathcal{M}_{\textbf{S}_a\textbf{S}_b}+ \sum_{\textbf{VB,S}}\mathcal{M}_{\textbf{VBS}}+\sum_{\textbf{VB}}\mathcal{M}_{\textbf{VB}}
+\mathcal{M}_{\gamma}
\Bigr),
\label{effective}
\end{equation}
where $\mathcal{M}_\gamma$ denote the box diagrams involving one photon and one $Z^\prime$ or one photon and one (pseudo)scalar.  The minus sign arises from the usual relation $\mathcal{H}_{\textbf{eff}} = - \mathcal{L}_{\textbf{eff}}$. 

Before providing the results we need to clarify some assumptions. First, once we are dealing with heavy degrees of freedom, the four-momenta for all external particles can be consider to be zero~\cite{CB}. Secondly, diagrams obtained by those in Figs.~\ref{fig7} and \ref{fig8} where the boson and quark lines are interchanged also do exist. We can show that the effective Hamiltonian derived from Fig.~\ref{fig7}(a) and Fig.~\ref{fig8} is the same for the two sort of diagrams, i.e., the action of rotating the internal lines does not change the final result to the amplitude and we have a 2 factor in our final result. Although the same is not true to those in Fig.~\ref{fig7}(b), we will anticipated a consequence for vacuum insertion and give only the terms that will not be zero.

Moreover, we will use the following notation
\begin{eqnarray}
&&[\gamma^\mu P_\zeta]_{jk}[\gamma^\nu P_\xi]_{jk}=[\bar{D}_j(p_3)\gamma^\mu P_\zeta D_k(p_4)][  \bar{D}_j(p_2)\gamma^\mu P_\xi D_k(p_1)],\nonumber \\&&
[P_\zeta]_{jk}[P_\xi]_{jk} =[\bar{D}_j(p_3) P_\zeta D_k(p_4)][\bar{D}_j(p_2) P_\xi D_k(p_1)],\quad\zeta,\;\xi=L,R.
\label{deflr}
\end{eqnarray} 
For the case of neutral kaons we have $k=1$ and $j=2$; $B_s$ are obtained when~$k=2,j=3$, and $B_d$ when $k=1, j=3$. See Figs.~\ref{fig7} - \ref{fig9} for the momenta assignment.

\subsection{Boxes with two scalar bosons }
\label{subsec:qs}

The first set of boxes are those in Fig.~\ref{fig7}(a), in which two charged or neutral scalars, denoted by $\textbf{S}_a\textbf{S}_b$ and two quarks, denoted $q_i$ and $q_l$, are involved. For fixed scalar indices, $a,b$, this large number of amplitudes are summarized as follows  
\begin{eqnarray}
i{\mathcal M}_{\textbf{S}_a\textbf{S}_b} &=&  2 \sum_{i,l} \biggl\{ (I^{\mu \nu}_{ \textbf{S}_a\textbf{S}_b})_{il}
\Bigl[\bar{D}_j(p_3)  \gamma _{\mu} \, \{ (\mathbf{K}_a)_{jl} ( \mathbf{K}_b)^*_{kl}  P_L + 
(\tilde{\mathbf{K}}_a)^*_{lj} (\tilde{\mathbf{K}}_b)_{lk} P_R \} D_k(p_4)\Bigr ] 
\nonumber \\
& & \Bigl[\bar{D}_j(p_2) \gamma _{\nu} \{ (\mathbf{K}_b)_{ji} ( \mathbf{K}_a)^*_{ki} P_L + (\tilde{\mathbf{K}}_b)^*_{ij} (\tilde{\mathbf{K}}_a)_{ik} P_R  \} D_k(p_1) \Bigr]  + \nonumber \\
& & + m_i m_l \, (I_{\textbf{S}_a\textbf{S}_b})_{il}  \Bigl[\bar{u}(p_3)   \{  (\mathbf{K}_a)_{jl} (\tilde{\mathbf{K}}_b)_{lk} P_R  +  
(\tilde{\mathbf{K}}_a)^*_{lj} ( \mathbf{K}_b)^*_{kl}  P_L \} v(p_4)\Bigr ] 
 \nonumber \\
& & \Bigl[\bar{v}(p_2) \{ (\mathbf{K}_b)_{ji} (\tilde{\mathbf{K}}_a)_{ik} P_R  + 
(\tilde{\mathbf{K}}_b)^*_{ij} ( \mathbf{K}_a)^*_{ki} P_L \} u(p_1) \Bigr] \biggr\},
\label{ampss}
\end{eqnarray}
where $\mathbf{K}_a,\tilde{\mathbf{K}}_a$ are the matrices related to the scalar boson $\mathbf{S}_a$ and run over the matrices showing in Eqs.~(\ref{a1e2}) - (\ref{a1e4}).
The indices $j,k$ relates to $K$ , $B_d$ or $B_s$ mesons, according to the convention discussed below Eq.~(\ref{deflr}). On the other hand, $i,l$ run over exotic or usual internal quarks. 

We can rewrite Eq.~(\ref{ampss}) by defining
\begin{eqnarray}
\mathbf{X}_{ab;jlkm}=(\mathbf{K}_a)_{jl}( \mathbf{K}_b)^*_{km},\quad
\tilde{\mathbf{X}}_{ab;jlkm}=( \tilde{\mathbf{K}}_a)_{jl}( \tilde{\mathbf{K}}_b)^*_{km},\quad 
\mathbf{Y}_{ab;jlkm}=( \mathbf{K}_a)_{jl}(\tilde{\mathbf{K}}_b)_{km}.
\label{defxys}
\end{eqnarray}
and the notation in Eq.~(\ref{deflr}) are now manifest:
\begin{eqnarray}
i{\mathcal M}_{\textbf{S}_a\textbf{S}_b} &=& 2 \sum_{i,l} \biggl\{ (I^{\mu \nu}_{\textbf{S}_a\textbf{S}_b})_{il}
\Bigl(
\mathbf{X}_{ab;jlkl}[\gamma_\mu P_L]_{jk}+\tilde{\mathbf{X}}_{ba;lklj}[\gamma_\mu P_R]_{jk}\Bigr)\nonumber \\&\cdot&\Bigl(\mathbf{X}_{ba;jiki} [\gamma_\nu P_L]_{jk}+\tilde{\mathbf{X}}_{ab;ijik} [\gamma_\nu P_R]_{jk}\Bigl)\nonumber \\&+&
m_i m_l\,(I_{\textbf{S}_a\textbf{S}_b})_{il}
\Bigl( \mathbf{Y}_{ab;jllk}[P_R]_{jk}+ \mathbf{Y}^*_{ba;kllj}[P_L]_{jk}\Bigr)
\nonumber \\ &\cdot& \Bigl(\mathbf{Y}_{ba;jiik}[P_R]_{jk}
+\mathbf{Y}^*_{ab;kiij}[P_L]_{jk}
\Bigr)\biggr\},
\label{ampa}
\end{eqnarray}
where $(I^{\mu\nu}_{\textbf{S}_a\textbf{S}_b})_{il}$ and $(I_{\textbf{S}_a\textbf{S}_b})_{il} $ denote the integrals
\begin{eqnarray}
(I^{\mu \nu}_{\textbf{S}_a\textbf{S}_b})_{il} &=&  \int \frac{d^4 k}{(2 \pi)^4} \frac{k^{\mu} k^{\nu}}{(k^2 - m_i^2)(k^2 - m_l^2) (k^2-m^2_{\textbf{S}_a})(k^2-m^2_{\textbf{S}_b})}, \nonumber \\ 
(I_{\textbf{S}_a\textbf{S}_b})_{il} &=&  \int \frac{d^4 k}{(2 \pi)^4} \frac{1}{(k^2 - m_i^2)(k^2 - m_l^2) (k^2-m^2_{\textbf{S}_a})(k^2-m^2_{\textbf{S}_b})}.
\label{int}
\end{eqnarray}

If in Eq.~(\ref{ampa}), $\textbf{S}_a = h^0$ and $\textbf{S}_b = A^0$ we have an overall minus sign. We recall that the complete amplitude for two scalars (physical or Goldstone, charged or neutral) and two quarks (known or exotic) will be obtained by summing over $\textbf{S}_a\textbf{S}_b$.

The integrals written in Eq.~(\ref{int}) define new Inami-Lim functions~\cite{Inami:1980fz} and the complete results, after the integrations being performed, are presented in Appendix~\ref{sec:integrals}.

\subsection{Boxes with one vector boson and one (pseudo)scalar}
\label{subsec:qvs}

When a vector and a scalar bosons are in the box, see Fig.~\ref{fig7}(b), the amplitude is written as
\begin{eqnarray}
i\mathcal{M}_{\textbf{VBS}_a} &=&  g^2 \sum_{i,l} \Bigl\{ 2 (I^{\mu\nu}_{\textbf{VBS}_a})_{il} \, (\mathbf{V})_{ji}  (\mathbf{V})^*_{kl} 
[\gamma _{\mu} P _L]_{jk} 
\{  \mathbf{X}_{aa;jlki} [\gamma _{\nu} P_L]_{jk}   + 
\tilde{\mathbf{X}}_{aa;iklj}[\gamma_\nu P_R]_{jk}   \} + \nonumber \\
& & +
 m_i m_l \, (I_{\textbf{VBS}_a})_{il} \mathbf{X}_{aa;jlki} [\gamma _{\mu} P_L]_{jk} [\gamma ^{\mu} P_L]_{jk}  \nonumber \\ 
 & & + g_{\mu\nu}(I^{\mu\nu}_{\textbf{VBS}_a})_{il} \tilde{\mathbf{X}}_{aa;iklj}[P_R]_{jk}[P_L]_{jk}
\Bigr\}.
\label{ampb} 
\end{eqnarray}
The matrix $\mathbf{V}$ depends on the vector boson in the box and the integral $I^{\mu\nu}_{\textbf{VBS}_a}$ is defined according to (\ref{int}) by replacing one of the scalar mass for one of the vector boson. There is a 2 factor that takes into account $\mathcal{M}_{\textbf{VBS}_a} = \mathcal{M}_{\textbf{S}_a\textbf{VB}}$. Again, if $\textbf{VB} = Z^\prime$ and $\textbf{S}_a = A^0$ we have an overall minus sign. We note that there is no sum in $\textbf{VB}$ and $\textbf{S}_a$.

\subsection{Boxes with two heavy vector bosons}
\label{subsec:qvb}

Here for massive vector bosons we consider $Z^\prime_\mu,V^\pm_\mu$ and $U^{\pm\pm}_\mu$. The typical box is shown in Fig.~\ref{fig8}.  For two fixed vector bosons we have
\begin{eqnarray}
i\mathcal{M}_{\textbf{VB}} &=& 2 g^4  \sum _{i, l} \,(I^{\mu \nu}_{\textbf{VB}})_{il} (\mathbf{V})_{jl}(\mathbf{V})^*_{kl}(\mathbf{V})_{ji}(\mathbf{V})^*_{ki} 
\,    [\gamma_{\mu} P_L]_{jk}[\gamma_{\nu} P_L]_{jk} + \nonumber \\
& & + g^4 (I^{\mu \nu}_{Z^\prime})_{jk} (K_L^D)^2_{jk}(K_R^D)_{kk}(K_R^D)_{jj}
\,   \bigl( [\gamma_{\mu} P_L]_{jk}[\gamma_{\nu} P_R]_{jk} + 
[\gamma_{\mu} P_R]_{jk}[\gamma_{\nu} P_L]_{jk} \bigr)
\label{ampc}  
\end{eqnarray}
and again $(I^{\mu\nu}_{\textbf{VB}})_{il}$ is defined as in (\ref{int}) with the vector masses instead of the scalar masses. The matrix $\mathbf{V}$ will be, for example, $V^D_L$ when one $V^\pm_\mu$ or $U^{\pm\pm}_\mu$ appear in the box. When $\textbf{VB} = Z^\prime$ the matrix $\mathbf{V}$ must be only $K_L^D$, since we have already included the constant right-handed part of the vertex. 
The matrix $K^D_R$ is proportional to the unit matrix, as shown in Subsec.~\ref{subsec:cc}. As before, we have included a 2 factor for the rotated diagram. 

Examples of the amplitudes in Eqs.~(\ref{ampa}), (\ref{ampb}) and (\ref{ampc}) are shown in the Appendix~\ref{sec:examples} for some selected cases. 

\subsection{Boxes with photon}
\label{subsec:photon}

\begin{center}
   	\begin{figure}[!ht]
   		\subfloat{\includegraphics[width=7cm]{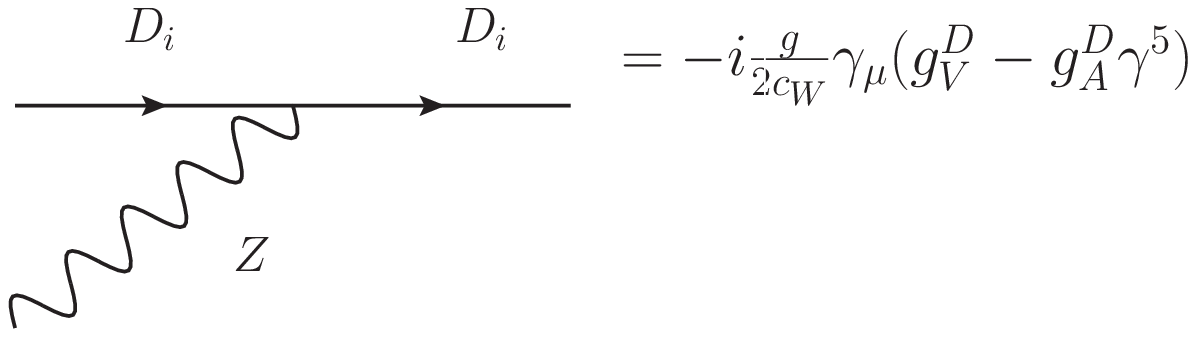}} \ \ \
   		\subfloat{\includegraphics[width=5cm]{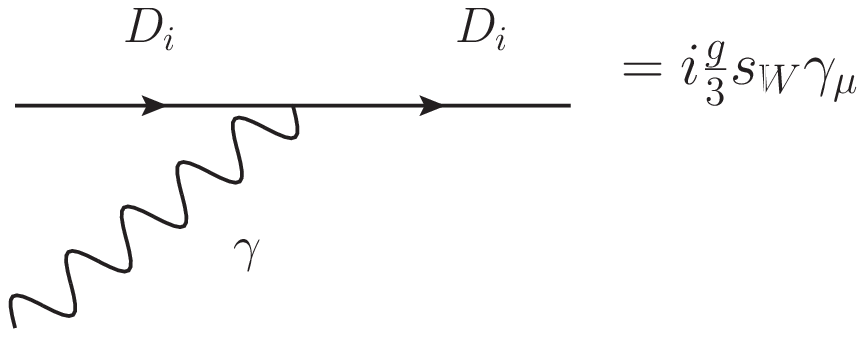}} 
   		\caption{Interaction vertices of $Z$ and photon, where $g_V^D = -\frac{1}{2} + \frac{2}{3}s_W^2$ and $g_A^D = -\frac{1}{2}$.}\label{fig51}
   	\end{figure}
   \end{center}

\begin{center}
 	\begin{figure}[H]
 		\subfloat[]{\includegraphics[width=7cm]{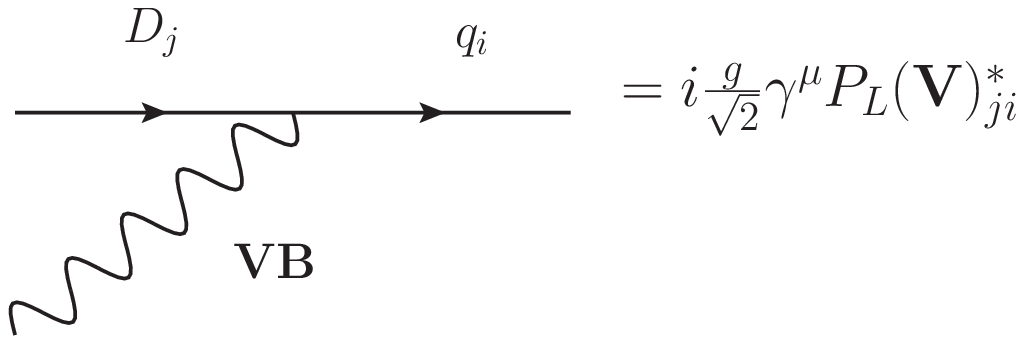}} \ \ \ 
 		\subfloat[]{\includegraphics[width=7cm]{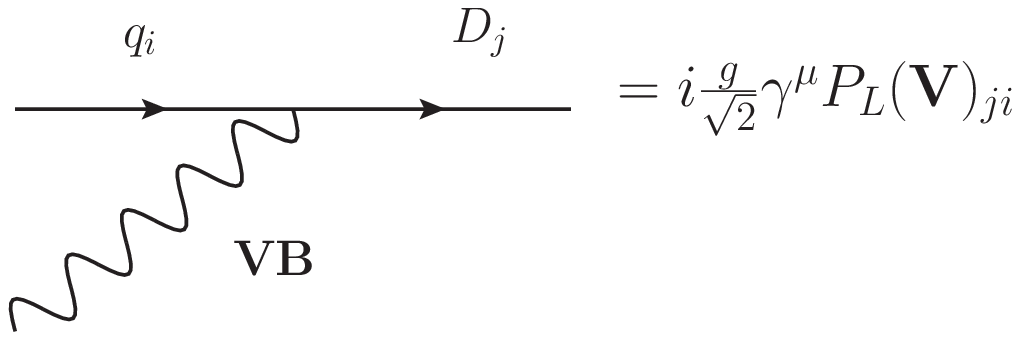}}
 		\caption{The generic vertices of quarks and vector bosons.}\label{fig6}
 	\end{figure}
 	
 \end{center}

\begin{center}
\begin{figure}[h]
\subfloat[]{\includegraphics[width=8cm]{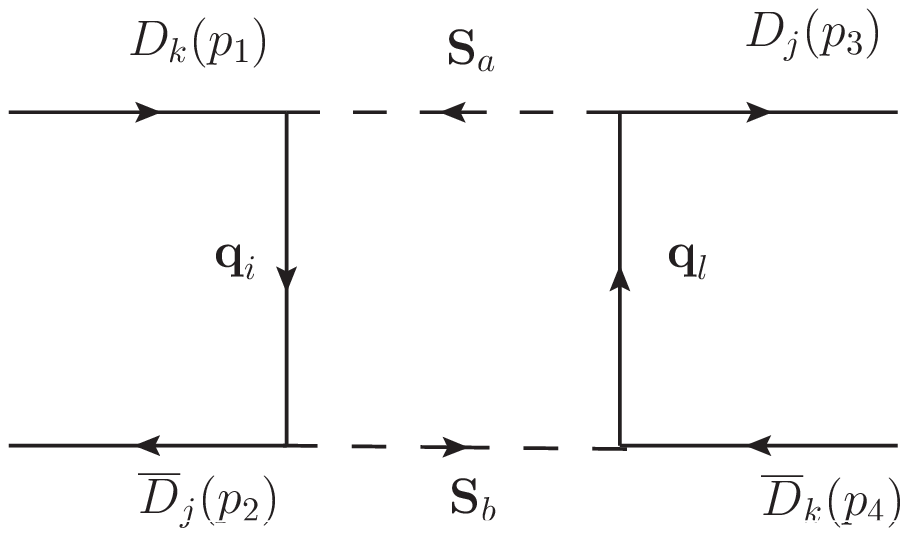}}
\subfloat[]{\includegraphics[width=8cm]{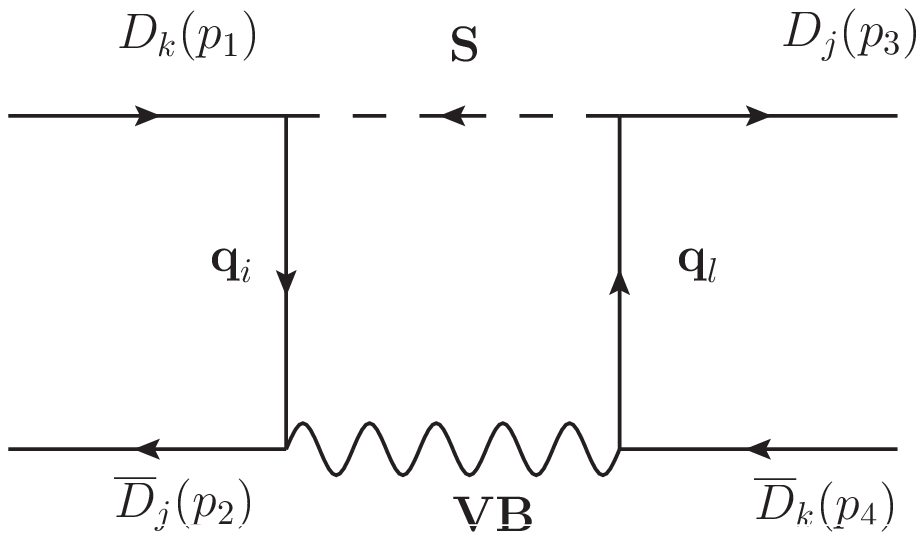}}
\caption{The general diagrams with (a) two scalars $\textbf{S}_a \textbf{S}_b$  and (b) with one vector boson and one scalar $\textbf{VBS}$.}\label{fig7}
\end{figure}
\end{center}

\begin{figure}[ht]
 \includegraphics[scale=0.8]{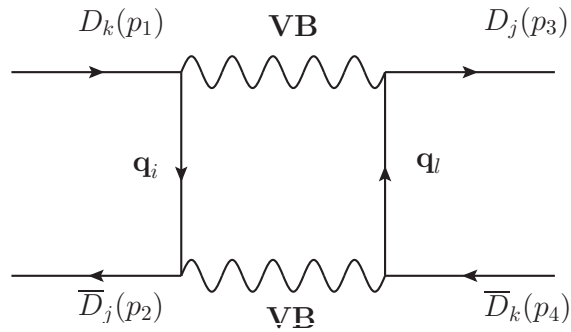}
 \caption{The general diagram with vector bosons, $\textbf{VB}$.} \label{fig8}
 \end{figure}

\begin{center}
 \begin{figure}[h]
 \subfloat[]{\includegraphics[width=8cm]{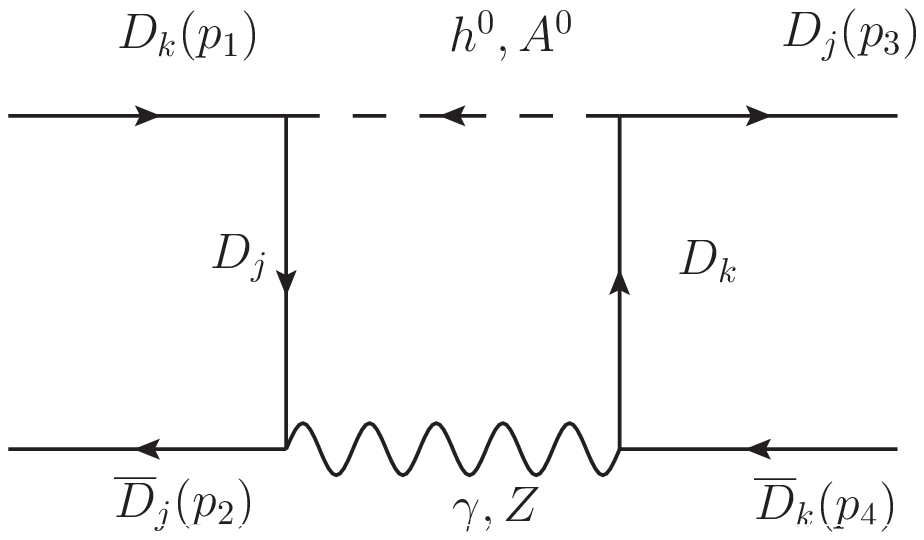}}
 \subfloat[]{\includegraphics[width=8cm]{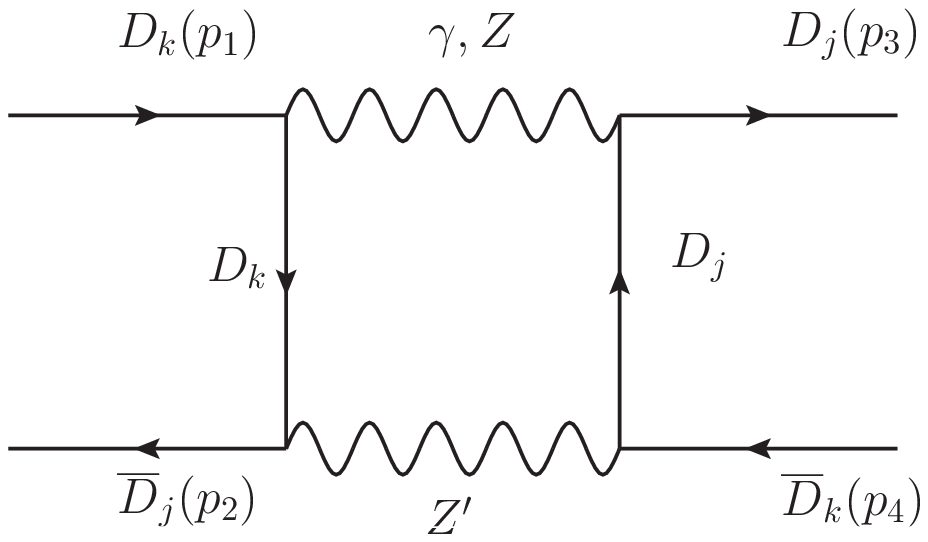}}
 \caption{The diagrams with photons (a) and scalars and (b) with $Z'$.}\label{fig9}
 \end{figure}
 \end{center}

The vertices in this kind of diagram are the usual ones, see Fig.~\ref{fig51} and Fig.~\ref{fig9}. Most of the bilinear are cancelled out by symmetry after vacuum insertion. Moreover, some terms are negligible because are proportional to the ratio $\frac{m_i m_j}{m_{331}^4}$, with $m_{i,j}$ the mass of an $d$-type quark and $m_{331}$ the mass of $Z^\prime$ or $h^0, A^0$. The amplitude is
\begin{eqnarray}
i\mathcal{M}_{\gamma} &=& \mathcal{M}_{\gamma Z^\prime} +\mathcal{M}_{\gamma h}-\mathcal{M}_{\gamma A} ,
\label{ampphot}  
\end{eqnarray}
where 
\begin{eqnarray}
&& \mathcal{M}_{\gamma Z^\prime}=\frac{4 g^4 s_W^2}{9} \,(I^{\mu \nu}_{\gamma Z'})_{jk} (K^D_L)^2_{jk}
\,    [\gamma_{\mu} P_L]_{jk}[\gamma_{\nu} P_L]_{jk},
\nonumber \\&& 
\mathcal{M}_{\gamma h}=\frac{g^2 s_W^2}{9} 
\{ g_{\mu \nu} (I^{\mu \nu}_{\gamma h})_{jk} 
[(\mathcal{K}^D_{h})_{jk} P_R + (\mathcal{K}^D_{h})_{kj}^* P_L]_{jk}[(\mathcal{K}^D_{h})_{jk} P_R + (\mathcal{K}^D_{h})^*_{kj} P_L]_{jk}  \},\nonumber \\&&
\mathcal{M}_{\gamma A}=\frac{g^2 s_W^2}{9} 
\{ g_{\mu \nu} (I^{\mu \nu}_{\gamma A})_{jk} 
[(\mathcal{K}^D_{A})_{jk} P_R + (\mathcal{K}^D_{A})_{kj}^* P_L]_{jk}[(\mathcal{K}^D_{A})_{jk} P_R + (\mathcal{K}^D_{A})^*_{kj} P_L]_{jk}  \},
\label{ghga}
\end{eqnarray}
and the indexes $j,k$ are fixed depending of the meson considered.  With the matrices in Eq.~ (\ref{a1e5}) and the input parameters in (\ref{input}) below, there is a negative interference between $\mathcal{M}_{\gamma h}$ and $\mathcal{M}_{\gamma A}$, hence it is much smaller than $\mathcal{M}_{\gamma Z^\prime}$.

\subsection{Boxes with one Z}
\label{subsec:Z}

In the SM limit of 331HL the Z boson preserve flavors \cite{Machado:2013jca}. Despite this feature, we can also have a few box diagrams with scalars or $Z^\prime$, just as the previous case. The amplitude is
\begin{eqnarray}
i\mathcal{M}_{Z} &=& \mathcal{M}_{ZZ^\prime} +\mathcal{M}_{Zh}-\mathcal{M}_{ZA}, 
\label{ampZ}  
\end{eqnarray}
where
\begin{eqnarray}
&& \mathcal{M}_{ZZ^\prime}=\frac{g^4}{c_W ^2} f_1(s_W ^2) \,(I^{\mu \nu}_{Z Z^\prime})_{jk} (K^D_L)^2_{jk}
\,    [\gamma_{\mu} P_L]_{jk}[\gamma_{\nu} P_L]_{jk},
\nonumber \\&& \mathcal{M}_{Zh}=2\frac{g^2}{c_W^2} f_2(s_W ^2) 
g_{\mu \nu} (I^{\mu \nu}_{Z h})_{jk} 
(\mathcal{K}^D_{h})_{jk}(\mathcal{K}^D_{h})_{kj}^*[P_R]_{jk}[P_L]_{jk},\nonumber \\ &&
\mathcal{M}_{ZA}=2\frac{g^2}{c_W^2} f_2(s_W ^2) 
g_{\mu \nu} (I^{\mu \nu}_{Z A})_{jk} 
(\mathcal{K}^D_{A})_{jk}(\mathcal{K}^D_{A})_{kj}^*[P_R]_{jk}[P_L]_{jk}
\label{fhfa}
\end{eqnarray}
with
\begin{equation} 
f_1(s_W^2) = (1-\frac{1}{3}s_W^2)^2 + (1 - \frac{2}{3}s_W^2),  \quad\ f_2(s_W^2) = (1-\frac{1}{3}s_W^2)^2 ,
\label{fZ}
\end{equation} 
and we are already considering the rotated diagram.

\section{Mass difference in the pseudoscalar mesons}
 \label{sec:mesons}
 
 We assume that the boxes involving only the SM particles are well-known. Then, the main purpose of this work is to verify in which realistic scenario the extra contributions coming from the new 331HL particles fulfill the requirement
 \begin{equation}
 \Delta m_M\vert^{full}_{331}-\Delta m_M\vert^{tree}_{331}-\Delta m_M\vert_{SM}=\Delta m_M\vert^{boxes} _{331} <10^{-15}, 10^{-13},10^{-11}\quad \textrm{GeV},
 \label{true}
 \end{equation}
 for $M = K, B_d, B_s$, respectively. The conditions above are enough to put lower limit on the mass of the extra particles in the model.
We will consider the amplitude $M^0\to \bar{M}^0$ (where $M^0=D_k\overline{D}_j$) arising such that, as usual,
 \begin{equation}
 \Delta m_M = 2 \textrm{Re} \langle \overline{M}^0 \vert \mathcal{H}_{\textbf{eff}}^{\Delta F = 2} \vert M^0 \rangle,
 \label{deltam}
 \end{equation}
 where $\mathcal{H}_{\textbf{eff}}^{\Delta F = 2}$ is the full effective Hamiltonian given by (\ref{effective}).
 
In order to obtain our final results we must be supported by some hints~\cite{Machado:2013jca} and choose possible values to the remaining free parameters:
\begin{equation} 
\begin{array}{ccc}
m_J = g_J\frac{v_\chi}{\sqrt2},& a_7 = a_8 = a_9 = 2, &   m_{h^0} = 125 ~\text{GeV},\\
g_J = 3.07  &  v_{\rho} = 54 ~\text{GeV},&  v_{\eta} = 240 ~\text{GeV},\\	
U_{\eta _1} = 0.1, & \qquad U_{\rho_1} = 0.42, & \\		
\end{array}
\label{input}
\end{equation} 
where $v_{\rho}, v_{\eta}$ are the vacuum expectation values for the scalars $\rho, \eta$, respectively, and we have assumed $m_{j_1} \approx m_{j_2}$. 
The Appendix~\ref{sec:scalarshl} introduces the constants $a_7, a_8, a_9$ from the scalar potential. $U_{\rho(\eta) 1}$ are matrix elements of the projection over the 125 GeV Higgs scalar of Re$\rho^0$ and Re$\eta^0$, respectively (In the case of m331 these matrices would be parametrized by general orthogonal matrices). As we said below Eq.~(\ref{esc2}), the choice of $U_{\rho1}=0.42$ means that the $\rho^0$ has the largest projection on the 125 GeV SM-like Higgs. 

We note that due to the relation between the mass and symmetry eigenstate of quarks $J_{1,2}$ in Eq.~(\ref{jotas}) the amplitudes involving two $j_{1,2}$ are proportional to $\cos^2\theta^2\sin^2\theta(m^2_{j_1}-m^2_{j_2})$. Hence,  there is a GIM-like mechanism in that sector and, under the conditions above, these contributions are negligible so that they do not impose strong constraints on the masses of $j_1$ and $j_2$. 

The gauge boson masses depend on $v_\chi$  and the scalar ones depend mainly on $v_\chi$ and $\alpha$, the latter one is an intermediate scale in the scalar potential. The mass dependence on these constants in the scalar sector are also presented in Appendix~\ref{sec:scalarshl}. Furthermore, on the gauge bosons we have

\begin{equation} 
\begin{array}{cc}
m^2_U \approx \frac{g^2}{4}(v^2_\rho + v^2_\chi), & \qquad\ m^2_V \approx \frac{g^2}{4}(v^2_\eta + v^2_\chi), \\
\end{array}
\label{input2}
\end{equation} 
and
\begin{equation}\label{zmass}
m^2_{Z'} \approx \frac{g^2}{2 c^2_W}\frac{(1-s_W^2)(4-\bar{v}_W)+s_W^4(4-\bar{v}^2_W)}{1-4s^2_W},		
\end{equation}  
where $\bar{v}_W \approx \frac{\sqrt{v^2_\rho +v^2_\eta}}{v_\chi}$. We have shown these masses as functions of $v_\chi$ in Fig.~\ref{fig10} - Fig.~\ref{fig11} and, from Eq.~(\ref{zmass}), $v_\chi$ must be greater than 67.3 GeV. Once we have performed the sum in Eq.~(\ref{effective}) and put it in Eq.~(\ref{deltam}), the $\Delta m_M$ could be obtained by using the matrix elements in Appendix \ref{sec:vi}. 

The final results are plotted in Fig.~\ref{fig12} - Fig.~\ref{fig14}. The respective upper limit in Eq.~(\ref{true}) to the three neutral mesons is reached, simultaneously,  under the condition $v_\chi > 820$ GeV, which is depicted by the vertical red line. Converted to the masses, we have (in TeV)
\begin{equation} 
\begin{array}{ccc}
m_U > 0.27, & \qquad m_V > 0.28, & \qquad m_{Z'} > 2.4, \\
m_{Y_1} > 20.1, & \qquad m_{Y_2} > 4.7, & \qquad \\
m_{Y^{++}} > 19.7, & \qquad m_{A^0} > 20.2. & \\
\end{array},
\label{result}
\end{equation}
The trilinear interactions in the scalar potential still depends on some constant $\alpha$ that we have chosen equals to $150$ TeV. the trilinear interactions in the scalarpotential.

We do not have to concern about negative values in Fig.~\ref{fig12}, since the conditions (\ref{true}) takes into account only the boxes. Our purpose is to find just an example in which the 1-loop corrections are more suppressed than the tree level amplitudes and also suppressed  with respect to the 1-loop SM. 
However, the results are obtained using the matrices $V^{U,D}_{L,R}$ from Ref.~\cite{Machado:2013jca}, reproduced in Appendix ~\ref{sec:matrices}.
These matrices are not unique and a distinct set could provide, at the same time, the correct quark masses and the CKM, which would imply another range of parameter and, finally, a different position to the vertical red line in Fig.~\ref{fig12} - Fig.~\ref{fig14}.

There are still some loop diagrams of the same order as the boxes, see Figs.~\ref{fig16} - \ref{fig17}. Nevertheless, these diagrams will be negligible and a result is presented in Appendix ~\ref{subsec:24}.
 
\section{Conclusions}
\label{sec:con}

Almost all the extensions of the ESM have a plenty of new free parameters. One important issue is to obtain the $SU(2)\otimes U(1)$ limit of a given extended model. This (trivial) limit occurs in the 3-3-1 models in the setting of infinite mass of the extra particles i.e., when $\vert v_\chi\vert\to\infty$, the $SU(3)_L\otimes U(1)$ and the
$SU(2)_L\otimes  U(1)_Y$ theories are equivalent. However, in Ref.~\cite{Dias:2006ns} it was discovery another (non-trivial) $SU(2)\otimes U(1)$ scenario to these models in such a way that there is no dependence on the VEV $v_\chi$ and  the values of the $v_\eta\approx54$ GeV and $v_\rho\approx240$ GeV are fixed. 

It was in this context that in Ref.~\cite{Machado:2013jca} numerical values of the $V^{U,D}_{L,R}$ matrices were obtained by fitting the quark masses and the CKM entries. As we said before they are not unique and it is possible to obtain other values for them giving the correct inputs. Hence, we would like to stress that our results depend on this non-trivial SM limit of the model \textit{and} on the numerical values obtained in \cite{Machado:2013jca} which assume that the sextet is not important to generate the lepton masses. Some authors follow the inverse way, accepting the $Z^\prime$ mass as an input and considering experimental data, say $C\!P$ violation and rare decays, in order to obtain the allowed values of the entries of $V^{U,D}_{L,R}$~\cite{Promberger:2007py}. 

In Eq.~(\ref{input}) we have used $g_J=3.07$ such that $m_J>1.78$ TeV for the mass of the exotic quark with electric charge of 5/3$\vert e \vert$. Lower limit of 800 GeV for the mass of this sort of quarks is obtained by experiment~\cite{Chatrchyan:2013wfa} assuming that they decay as $T\to W^+t$ and then $t\to W^+b$. However, in the m331 and 331HL models, the $J$-quark decays as $J\to tV^+,bU^{++}$. We note that in the former case the signature is two jets or missing energy, while in the latter case, the signature is two leptons with the same sign, $U^{++}\to l^+l^{\prime +}$. Similarly, a top quark with -4/3 decaying 
as $T\to bW^-$ was ruled out by CDF~\cite{Aaltonen:2010js}. In the present model nonetheless quarks with charge -4/3, denoted by $j_{1,2}$, decay as $j_{1,2}\to tU^{--},bV^-$.  In conclusion, the experimental lower limit does not apply, at least in a straightforward way, to our case.

Experimental searches for $W^\prime$ have also been performed searching for $W^\prime \to tb \to l\nu_lbb$ implying that $M_{W^\prime}>1.84$ TeV~\cite{Chatrchyan:2014koa}. The singly charged extra vector field in m331 and 331HL models decays, for instance, as $V^-\to j_{1,2}\bar{b}$.  Concerning the searching for $Z^\prime$, we note that this vector boson also has decays that are different from other models with this sort of neutral boson. Furthermore, the neutral (pseudo)scalars may interfere positively or negatively and have to be considered.
 
Although there are more than three hundreds of new box diagrams (with respect to their number in the SM) for meson mixing in the framework of 331HL model, we have classified them in four types. We have shown that the evaluation of these diagrams provides a powerful mechanism to test different scenarios to the free parameters of the model. 

\acknowledgments

The authors would like to thank E.C.F.S Fortes, A.C.B. Machado, G. De Conto and J. Montaño for the insightful comments. This work was fully (FCC) and partially (VP) supported by CNPq.

\newpage

\appendix

\section{Matrices}
\label{sec:matrices}

In the 331HL, there are several matrices in the flavour space. 
In this appendix we write them explicitly and also show the numerical values using the parameters that give the quark masses and the CKM matrix obtained in Ref.~\cite{Machado:2013jca}.

From Eq.~(\ref{q1}) the Yukawa interactions between quarks and charged scalar  in Eq.~(\ref{esc1}) involve the matrices from (\ref{a1e2}) to 
(\ref{a1e4}). Defining
\begin{eqnarray}
&&N_1= \frac{1}{\sqrt{1+\frac{|v_{\chi}|^2}{|v_{\eta}|^2}}}\approx\frac{|v_{\eta}|}{|v_{\chi}|} ,\quad N_2=\frac{1}{\sqrt{1+\frac{|v_{\eta}|^2}{|v_{\chi}|^2}}}\approx1,\quad 
N_3=\frac{1}{\sqrt{1+\frac{|v_{\chi}|^2}{|v_{\rho}|^2}}} \approx\frac{|v_{\rho}|}{|v_{\chi}|},\nonumber \\&& N_4=\frac{1}{\sqrt{1+\frac{|v_{\rho}|^2}{|v_{\chi}|^2}}}\approx1,\quad
N_5=\frac{1}{\sqrt{1+\frac{|v_{\rho}|^2}{|v_{\eta}|^2}}},\quad 
\label{a1e1}
\end{eqnarray}
we have
\begin{eqnarray}
\tilde{K}_1 &\!\!=\!\!& N_1\left( \begin{array}{ccc}
\tilde{G}_{11} & \tilde{G}_{12} & \tilde{G}_{13} \\ \tilde{G}_{21} & \tilde{G}_{22} & \tilde{G}_{23} \\ 0 & 0 & 0
\end{array}\right)
\left(V_R^D\right)^\dagger, \qquad 
K_1 = - \frac{\sqrt{2}}{|v_\chi|} N_2\left(V_L^D\right)
\begin{pmatrix}
m_{j_1} && 0 && 0 \\ 0 && m_{j_2} && 0 \\ 0 && 0 && 0
\end{pmatrix},
\nonumber  \\
\tilde{K}_2 &=& N_2 \begin{pmatrix}
\tilde{G}_{11} && \tilde{G}_{12} && \tilde{G}_{13} \\ \tilde{G}_{21} && \tilde{G}_{22} && \tilde{G}_{23} \\ 0 && 0 && 0
\end{pmatrix}
\left(V_R^D\right)^\dagger, \qquad
K_2 = \frac{\sqrt{2}}{|v_\chi|} N_1 \left(V_L^D\right)
\begin{pmatrix}
m_{j_1} && 0 && 0 \\ 0 && m_{j_2} && 0 \\ 0 && 0 && 0
\end{pmatrix}.
\label{a1e2}
\end{eqnarray}

\begin{eqnarray}
\tilde{K}_3 &=&  N_3  \begin{pmatrix}
0 && 0 && 0 \\ 0 && 0 && 0  \\ \tilde{F}_{31} && \tilde{F}_{32} && \tilde{F}_{33}
\end{pmatrix}
\left(V_R^D\right)^\dagger, \qquad
K_3 =  -N_4 
\begin{pmatrix}
0 && 0 && \left(V_L^D\right)_{13}  \\ 0 && 0 && \left(V_L^D\right)_{23}  \\ 0 && 0 && \left(V_L^D\right)_{33}
\end{pmatrix},
\nonumber \\
\tilde{K}_4 &=& N_4   \begin{pmatrix}
0 && 0 && 0 \\ 0 && 0 && 0  \\ \tilde{F}_{31} && \tilde{F}_{32} && \tilde{F}_{33}
\end{pmatrix}
\left(V_R^D\right)^\dagger, \nonumber \qquad
K_4 = N_3   
\begin{pmatrix}
0 && 0 && \left(V_L^D\right)_{13}  \\ 0 && 0 && \left(V_L^D\right)_{23}  \\ 0 && 0 && \left(V_L^D\right)_{33}
\end{pmatrix}. 
\label{a1e3}
\end{eqnarray}

Defining $\lambda = \frac{|v_{\rho}|}{|v_{\eta}|}$.
\begin{eqnarray}
K_5 &=&  N_5   \left(V_L^D\right)
\begin{pmatrix}
\lambda G_{11} &&  \lambda G_{12} &&  \lambda G_{13} \\
\lambda G_{21} &&  \lambda G_{22} &&  \lambda G_{23} \\
- F_{31} && - F_{32} && - F_{33}
\end{pmatrix} 
\left(V_R^U\right)^{\dagger} \approx  \begin{pmatrix}
-0.003  && 0.001 && -0.003 \\ 0.12 && -0.053 && 0.143 \\ -0.406  && 0.183  && -0.498 
\end{pmatrix},
\nonumber \\
K_6 &=&  N_5 \left(V_L^D\right)
\begin{pmatrix}
G_{11} &&  G_{12} && G_{13} \\
G_{21} &&  G_{22} &&  G_{23} \\
\lambda F_{31} && \lambda F_{32} && \lambda F_{33}
\end{pmatrix} 
\left(V_R^U\right)^{\dagger} \approx  \begin{pmatrix}
0.013  && -0.005 && 0.015  \\ -0.534 && 0.234  && -0.638  \\ 1.803 && -0.813 && 2.214 
\end{pmatrix},
\nonumber \\
\tilde{K}_5 &=&  - N_5 \left(V_L^U\right)
\begin{pmatrix}
\tilde{G}_{11} &&  \tilde{G}_{12} &&  \tilde{G}_{13} \\
\tilde{G}_{21} &&  \tilde{G}_{22} &&  \tilde{G}_{23} \\
\lambda \tilde{F}_{31} && \lambda \tilde{F}_{32} && \lambda \tilde{F}_{33}
\end{pmatrix} 
\left(V_R^D\right)^{\dagger}
\nonumber \\
&\approx & \begin{pmatrix}
2 \cdot 10^{-5} && 3 \cdot 10^{-5} &&  -8.9 \cdot 10^{-6} \\ 6.9 \cdot 10^{-5} &&  3.3 \cdot 10^{-4} && -3.2 \cdot 10^{-4} \\ -2.4 \cdot 10^{-5} && -3.8 \cdot 10^{-3} && 1.1 \cdot 10^{-2}
\end{pmatrix}, \nonumber \\
\tilde{K}_6 &=&  N_5 \left(V_L^U\right)
\begin{pmatrix}
- \lambda \tilde{G}_{11} && - \lambda \tilde{G}_{12} && - \lambda \tilde{G}_{13} \\
- \lambda \tilde{G}_{21} && - \lambda \tilde{G}_{22} && - \lambda \tilde{G}_{23} \\
\tilde{F}_{31} && \tilde{F}_{32} && \ \tilde{F}_{33}
\end{pmatrix} 
\left(V_R^D\right)^{\dagger} \nonumber \\
&\approx&  \begin{pmatrix}
1 \cdot 10^{-4} && 2 \cdot 10^{-4} && -1 \cdot 10^{-4} \\ -5.9 \cdot 10^{-6} && -5.6\cdot 10^{-5} && 6 \cdot 10^{-5}  \\ 5.46  \cdot 10^{-6} && 8.5 \cdot 10^{-4} && -2.4 \cdot 10^{-3}
\end{pmatrix}.
\label{a1e4}
\end{eqnarray}

In the matrices above we have used $G_{11}=1.08,G_{12}=2.97,G_{13}=0.09,G_{21}=0.0681,
G_{22}=0.2169,G_{23}=0.1\times10^{-2}$,
$F_{31}=9\times10^{-6},F_{32}=6\times10^{-6},F_{33}=1.2\times10^{-5}$, $\tilde{G}_{11}=0.0119,
\tilde{G}_{12}=6\times10^{-5},\tilde{G}_{13}=2.3\times10^{-5},\tilde{G}_{21}=(3.2 - 6.62)\times10^{-4},\tilde{G}_{22}=
2.13\times10^{-4},\tilde{G}_{23}=7\times10^{-5}$, $\tilde{F}_{31}=2.2\times10^{-4},
\tilde{F}_{32}=1.95\times10^{-4},\tilde{F}_{33}=1.312\times10^{-4}$. 
We emphasize that these parameters, as in any model with FCNC, may be not unique and it is possible that there exist other sets of values with which the quark masses and the CKM mixing matrices are obtained.

The matrices appearing in Eq.~(\ref{esc2}) are

\begin{eqnarray}
\mathcal{K}^D_{h}&\approx& 
\begin{pmatrix}
10^{-4} U_{\rho 1} - 10^{-6} U_{\eta 1} && 10^{-4} U_{\rho 1} - 10^{-5} U_{\eta 1} && 
-10^{-4} U_{\rho 1} + 10^{-5} U_{\eta 1} \\
10^{-6} U_{\rho 1} + 10^{-4} U_{\eta 1}&& 10^{-5} U_{\rho 1} + 10^{-3} U_{\eta 1}&& 
-10^{-6} U_{\rho 1} + 10^{-2} U_{\eta 1} \\
10^{-6} U_{\rho 1} - 10^{-5} U_{\eta 1} && 10^{-6} U_{\rho 1} - 10^{-3} U_{\eta 1} &&
 -10^{-6} U_{\rho 1} +0.011 U_{\eta 1}\end{pmatrix},
\nonumber \\
\mathcal{K}^D_{A}&\approx& 
\begin{pmatrix}
10^{-4} V_{\rho 1} - 10^{-6} V_{\eta 1} && 10^{-4} V_{\rho 1} - 10^{-5}  V_{\eta 1} && 
-10^{-4} V_{\rho 1} + 10^{-5} V_{\eta 1} \\
10^{-6}V_{\rho 1} + 10^{-4} V_{\eta 1}&& 10^{-5} V_{\rho 1} + 10^{-3}  V_{\eta 1}&& 
-10^{-6} V_{\rho 1} + 10^{-2}  V_{\eta 1} \\
10^{-6}V_{\rho 1} - 10^{-5}  V_{\eta 1} && 10^{-6}V_{\rho 1} - 10^{-3}  V_{\eta 1} &&
 -10^{-6} V_{\rho 1} +0.011  V_{\eta 1}
\end{pmatrix}.
\label{a1e5} 
\end{eqnarray}
where
\begin{equation} 
\begin{array}{cc}
V_{\eta 1} = \frac{1}{|v_\eta|}\left(\frac{1}{|v_\chi|^2}+\frac{1}{|v_\rho|^2}+\frac{1}{|v_\eta|^2}\right)^{-1/2}
, & \qquad 
V_{\rho 1} = \frac{1}{|v_\rho|}\left(\frac{1}{|v_\chi|^2}+\frac{1}{|v_\rho|^2}+\frac{1}{|v_\eta|^2}\right)^{-1/2}. \\
\end{array}
\end{equation}  

The matrix appearing in Eq.~(\ref{zprime}) is
\begin{equation}
K^D_L\approx  \frac{1}{\sqrt{2}c_W} 
\begin{pmatrix}
1.05154 && 0.00140  && 0.00826  \\
 0.00140 && -1.13082  && 5 \cdot 10^{-6}  \\
0.00826 && 5 \cdot 10^{-6} && - 1.13078
\end{pmatrix} \label{a1e6}
\end{equation}

The matrices  in the vector boson-quark interactions in Eq.~(\ref{lvb}) are given by:
\begin{eqnarray}
&& V^U_L\approx \left(\begin{array}{ccc}
-0.00032 & 0.07163&  -0.99743\\
0.00433& -0.99742  & -0.07163\\
0.99999& 0.00434& -0.00001\\
\end{array}\right), \\&&
 V^D_L\!\!\approx\!\!\left(\begin{array}{ccc}
0.00273 \to0.00562& (0.03\to0.03682)& -(0.99952\to0.99953)\\
-(0.19700\to0.22293)& -(0.97436\to0.97993) & -0.03052\\
0.97483\to0.98039& -(0.19708\to0.22291)& -(0.00415\to 0.00418)\nonumber \\
\end{array}\right).
\label{a1e7}
\end{eqnarray}

In the same way we obtain the $V^{U,D}_R$ matrices which will appear in the Yukawa interactions:
\begin{eqnarray}
&& V^U_R\approx \left(\begin{array}{ccc}
-0.45440 & 0.82278 &  -0.34139\\
0.13857& -0.31329 &-0.93949 \\
0.87996& 0.47421& -0.02834\\
\end{array}\right), \\&&
 V^D_R\!\!\approx\!\!\left(\begin{array}{ccc}
-(0.000178\to0.000185)& (0.005968 \to0.005984 )& -0.999982\\
-(0.32512\to0.32559)& -(0.94549\to0.94566) & -(0.00558\to0.00560)\\
0.94551\to0.94567& -(0.32511\to0.32558)& -(0.00211-0.00212 ) \nonumber \\
\end{array}\right).
\label{a1e8}
\end{eqnarray}

With this matrices the correct quark masses at the $Z$-pole and the CKM matrix were obtained in Ref.~\cite{Machado:2013jca}.

\section{Integrals}
\label{sec:integrals}

The integrals defined in Eq.~(\ref{int}) vary according the quarks and bosons in the internal lines. Thus, there are integrals with one, two and three variables. 

\begin{enumerate}
	
\item One variable
	
\begin{eqnarray}
I^{\mu \nu}_{A} (y_B) &=&  \int \frac{d^4 k}{(2 \pi)^4} \frac{k^{\mu} k^{\nu}}{(k^2 - m_A ^2)^2 (k^2-m^2_{B})^2} \label{Iuv1} \nonumber \\
&= &  - \frac{i g^{\mu  \nu}}{32 \pi ^2 m^2 _{A}} \frac{1 -y_B ^2 + 2x  \ln y_B}{2(y_B-1)^3} \\
I_{A}(y_B) &=&  \int \frac{d^4 k}{(2 \pi)^4} \frac{1}{(k^2 - m_A^2)^2 (k^2-m^2_B)^2} \label{I01} \nonumber \\
&= &    \frac{i}{16 \pi ^2 m^4 _A} \frac{2 -2y_B + (1+y_B) \ln y_B}{(y_B -1)^3}
\label{uma}
\end{eqnarray}
with 
\begin{equation}
y_B = \frac{m^2_B}{m^2_A}.
\end{equation}
	
This sort of functions will appear, for example, in diagrams with two quarks and two physical scalars.
	
\item Two variables
	
\begin{eqnarray}
I^{\mu \nu}_A (x_B, y_C) &=&  \int \frac{d^4 k}{(2 \pi)^4} \frac{k^{\mu} k^{\nu}}{(k^2-m^2_A)(k^2 - m^2_B)
(k^2 - m_C^2)^2} \label{Iuv2} \nonumber \\
&=&  - \frac{i g^{\mu  \nu}}{32 \pi ^2 m^2 _A}\,\frac{1}{2(y_C -1)^2 (y_C - x_B)^2 (x_B -1)} 
\nonumber \\&\cdot&\Biggl[ y_C (x_B -1)\bigl[ (y_C -1)(y_C - x_B) - (y_C +(y_C -2)x_B) \ln y_C \bigr] \nonumber \\ 
&+ &   (y_C -1)^2 x_B ^2 \ln x_B  \Biggr], \nonumber \\
I_A (x_B, y_C) &=&  \int \frac{d^4 k}{(2 \pi)^4} \frac{1}{(k^2-m^2_{A})(k^2 - m^2_B)(k^2 - m^2_C)^2 } \label{I02} \nonumber \\
&=& \frac{i}{16 \pi ^2 m^4 _A}\frac{1}{(y_C-1)^2(y_C-x_B)(x_B -1)} \Biggl[ (y_C^2 - x_B)(x_B-1)\ln y_C \nonumber \\
&- &  (y_C -1)\bigl[ (y_C -x_B)(x_B-1) + (y_C-1)x_B  \ln x_B \bigr] \Biggr], 
\label{duas}
\end{eqnarray}   
with $x_B = \frac{m^2 _B}{m^2 _A}$ e $y_C = \frac{m^2 _C}{m^2 _A}$.

\item Three variables 
	
Finally, we have the most general integral. As we have assumed  $m_{j_1} = m_{j_2}$, only diagrams with usual quarks will give rise to these sort of functions. Nevertheless, the scalar functions $I_A(x_B, x_C, x_D)$ will be suppressed by a $m_im_l/m^4_{331}$ factor, where $m_{331}$ is the typical mass of some new 331HL particles, always heavier than any SM quarks.
	
\begin{eqnarray}
I_A^{\mu \nu}(x_B, x_C, x_D) &=&  \int \frac{d^4 k}{(2 \pi)^4} \frac{k^{\mu} k^{\nu}}{(k^2 - m_A ^2)(k^2-m^2_B)(k^2 - m^2 _{C})(k^2 - m^2_{D})}  \nonumber\\
&=& - \frac{i g^{\mu  \nu}}{32 \pi ^2 m^2_{A}}\frac{1}{f(x_A,x_B,x_C)}\Biggl[ 
x_B^2 (x_C-1)(x_C-x_D)(x_D-1)\ln x_B\nonumber \\& -& x_C^2(x_B-1)(x_B-x_D)(x_D-1)\ln x_C   \nonumber \\
&+ &  x_D^2(x_B-1)(x_B-x_C)(x_C-1)\ln x_D \Biggr], 
\label{tres}
\end{eqnarray}   
where we have defined 
\begin{equation}
f(x_A,x_B,x_C)=2(x_B-1)(x_B-x_C)(x_C-1)(x_B-x_D)(x_C-x_D)(x_D-1),
\label{efe}
\end{equation}
with $x_{\alpha} = \frac{m^2 _{\alpha}}{m^2 _{A}}$, $\alpha = B,C,D$.	
	
\end{enumerate}

We can obtain the previous functions taking the limit

\begin{eqnarray}
\lim _{x_C \rightarrow x_D} I_A^{\mu \nu}(x_B, x_C, x_D) = I^{\mu \nu}_A (x_B, y_C) \hspace{2cm} 
\lim _{x_B \rightarrow 1} I^{\mu \nu}_A (x_B, y_C) = I^{\mu \nu}_A (y_C)
\end{eqnarray}

\newpage

\section{The scalar sector for the 331HL model}
\label{sec:scalarshl}

The most general scalar potential, invariant under CP transformations, for the scalars is:
\begin{eqnarray}
\label{potencial_escalar}
V(\chi,\eta,\rho)&=& \sum_{i} \mu^2_{i} \phi^\dagger_i \phi_i
+\sum_{i=1}^3 a_i(\phi^{\dagger}_i\phi_i)^2 +\sum_{m=4}^6 \sum _{\substack{i,j={1}\\ i>j}}^3 a_m(\phi_i^\dagger \phi_i)(\phi_j^\dagger \phi_j)
\nonumber \\&+&
\sum_{n=7}^9 \sum _{\substack{i,j={1}\\ i>j}}^3 a_n (\phi^{\dagger}_i\phi_j)(\phi^\dagger_j\phi_i) +( \alpha \,\epsilon_{ijk}\chi_{i}\rho_{j}\eta_{k}+h.c.) ,
\label{ace1}
\end{eqnarray}
where we have used $\phi_1=\chi,\phi_2=\eta$ and $\phi_3=\rho$, except in the trilinear term.

The mass spectra of the model has been obtained in Ref.~\cite{DeConto:2014fza}. Here we will summarized it. 
The doubly charged scalars $\rho^{++},\chi^{++}$ are related to the mass eigenstates as follows: 
\begin{equation}
\left(\begin{array}{c}
\rho^{++} \\ \chi^{++}
\end{array}\right)=\frac{1}{\sqrt{1+\frac{|v_\chi|^2}{|v_\rho|^2}}}
\left(\begin{array}{cc}
1 & \frac{|v_\chi|}{|v_\rho|}e^{-i\theta_\chi} \\ -\frac{|v_\chi|}{|v_\rho|}e^{i\theta_\chi} & 1
\end{array}\right)
\left(\begin{array}{c}
G^{++}_U \\ Y^{++}
\end{array}\right),
\label{ace2}
\end{equation} 
and the masses are
\begin{equation} 
m^2_{G^{++}_U}=0,\quad
m^2_{Y^{++}}=A\left(\frac{1}{|v_\rho|^2}+\frac{1}{|v_\chi|^2}\right)+\frac{a_8}{2}\left(|v_\chi|^2+|v_\rho|^2\right),
\label{ace3}
\end{equation}
where $ A= |v_\chi||v_\eta||v_\rho||\alpha|/\sqrt{2}$.

The singly charged scalars carrying no lepton number $\eta^+_1,\rho^+$ are related to the mass eigenstates as follows:
\begin{equation}
\left(\begin{array}{c}
\eta_1^+ \\ \rho^+
\end{array}\right)=\frac{1}{\sqrt{1+\frac{|v_\rho|^2}{|v_\eta|^2}}}
\left(\begin{array}{cc}
1 & \frac{|v_\rho|}{|v_\eta|} \\ -\frac{|v_\rho|}{|v_\eta|} & 1
\end{array}\right)
\left(\begin{array}{c}
G^+_W \\ Y_1^+
\end{array}\right),
\label{ace4}
\end{equation} 

with masses
\begin{equation} 
m^2_{G^+_W}=0,\quad
m^2_{Y_1^+}=A\left(\frac{1}{|v_\rho|^2}+\frac{1}{|v_\eta|^2}\right)+
\frac{a_9}{2}\left(|v_\eta|^2+|v_\rho|^2\right).
\label{ace5}
\end{equation}

In the singly charged scalars carrying lepton number $\eta^+_2,\chi^+$ we have
\begin{equation}
\left(\begin{array}{c}
\eta_2^+ \\ \chi^+
\end{array}\right)=\frac{1}{\sqrt{1+\frac{|v_\chi|^2}{|v_\eta|^2}}}
\left(\begin{array}{cc}
1 & \frac{|v_\chi|}{|v_\eta|}e^{i\theta_\chi} \\ -\frac{|v_\chi|}{|v_\eta|}e^{-i\theta_\chi} & 1
\end{array}\right)
\left(\begin{array}{c}
G^+_V \\ Y_2^+
\end{array}\right),
\label{ace6}
\end{equation} 
with masses
\begin{equation} 
m^2_{G^+_V}=0,\quad
m^{2}_{Y_2^+}=A\left(\frac{1}{|v_\chi|^2}+\frac{1}{|v_\eta|^2}\right)+
\frac{a_7}{2}\left(|v_\eta|^2+|v_\chi|^2\right).
\label{ace7} 
\end{equation}

Finally, in the neutral $C\!P$-odd scalars we have
\begin{equation}
\left(\begin{array}{c}
I^0_\eta \\ I^0_\rho \\ I^0_\chi
\end{array}\right)=
\left(\begin{array}{ccc}
\frac{N_a}{|v_\chi|} & - \frac{N_b|v_\eta||v_\chi|}{|v_\rho|(|v_\eta|^2+|v_\chi|^2)} & \frac{N_c}{|v_\eta|} \\
0 & \frac{N_b}{|v_\chi|} & \frac{N_c}{|v_\rho|} \\
-\frac{N_a}{|v_\eta|} & - \frac{N_b|v_\eta|^2}{|v_\rho|(|v_\eta|^2+|v_\chi|^2)} & \frac{N_c}{|v_\chi|}
\end{array}\right)
\left(\begin{array}{c}
G^0_1 \\ G^0_2 \\ A^0
\end{array}\right),
\label{ace8}
\end{equation} 
with the respective masses
\begin{equation} 
m^2_{G_1^0}=m^2_{G_2^0}=0,\quad
m^2_{A^0}=A\left(\frac{1}{|v_\chi|^2}+\frac{1}{|v_\rho|^2}+\frac{1}{|v_\eta|^2}\right).
\label{ace9}
\end{equation}

Above, we have defined
\begin{eqnarray}
&&N_a=\left(\frac{1}{|v_\chi|^2}+\frac{1}{|v_\eta|^2}\right)^{-1/2},
\quad
N_b=\left(
\frac{1}{|v_\chi|^2}+\frac{|v_\eta|^2}{|v_\rho|^2(|v_\eta|^2+|v_\chi|^2)}\right)^{-1/2},
\nonumber \\&&
N_c=\left(\frac{1}{|v_\chi|^2}+\frac{1}{|v_\rho|^2}+\frac{1}{|v_\eta|^2}\right)^{-1/2}.
\end{eqnarray}

For the $C\!P$-even scalars the mass matrix is real and symmetric and
we know that it can be diagonalized by an orthogonal matrix. Therefore:  $X^0_\psi=\sum_i U^H_{\psi i}H^0_i$, where $\psi=\chi,\eta,\rho$, $i=1,2,3$, $H^0_i$ are the mass eigenstates and $U^H$ is an orthogonal matrix.

We observe that in the minimal 3-3-1 model with the scalar sextet, there are four doubly charged scalars, six singly charged scalars which and five neutral scalars. Hence, the mass matrices are diagonalized by $4\times4$, $6\times6$ and $5\times5$ orthogonal matrices, respectively. If the lepton number is not explicitly violated in the potential, the singly charged scalar mass matrix is the direct sum of two $3\times3$ blocks, one in the $\eta^-_1,\rho^-,\sigma^-_2$ sector and another in the $\eta^-_2,\chi^-,\sigma^-_1$. It means that our results do not applied, at least in a straightforward way, to that model.

\section{Vacuum insertion approximation}
\label{sec:vi}

The final result to the mass difference of neutral mesons depends on the expectation value of the product of two bilinear as those shown in Eq.~(\ref{deflr}). In vacuum insertion approximation these matrix elements, as presented in \cite{CB}, can be given by: 
\begin{eqnarray}
&&\langle \bar{M} ^0 | (\bar{a}  \gamma^{\mu} P_L  q)(\bar{a}  \gamma_{\mu} P_L  q) | M^0 \rangle 
=\frac{1}{3} f_{M}^2 m_{M}, \\&&
\langle \bar{M}^0 | (\bar{a}  \gamma^{\mu} P_R  q)(\bar{a}  \gamma_{\mu} P_R  q) | M^0 \rangle 
=\frac{1}{3} f_{M}^2 m_{M}, \\&& 
\langle \bar{M}^0 | (\bar{a}  \gamma^{\mu} P_L  q)(\bar{a}  \gamma_{\mu} P_R  q) | M^0 \rangle 
=\frac{1}{6} f_{M}^2 \frac{m^3_{M}}{(m_q + m_a)^2} - \frac{1}{4} f_{M}^2 m_{M}, \\&&
\langle \bar{M} ^0 | (\bar{a}  \gamma^{\mu} P_R q)(\bar{a}  \gamma_{\mu} P_L  q) | M^0 \rangle 
=\frac{1}{6} f_{M}^2 \frac{m^3_{M}}{(m_q + m_a)^2} - \frac{1}{4} f_{M}^2 m_{M},  \\&&
\langle \bar{M} ^0 | (\bar{a} P_L  q)(\bar{a} P_L  q) | M^0 \rangle 
= \frac{5}{24} f_{M}^2 \frac{m^3_{M}}{(m_q + m_a)^2},
 \\&&
\langle \bar{M} ^0 | (\bar{a}  P_R  q)(\bar{a}  P_R  q) | \aleph ^0 \rangle 
= \frac{5}{24} f_{M}^2 \frac{m^3_{M}}{(m_q + m_a)^2},
 \\&&
\langle \bar{M} ^0 | (\bar{a} P_L  q)(\bar{a} P_R  q) | M^0 \rangle 
= \frac{1}{24} f_{M}^2 m_{M} - \frac{1}{4} f_{M}^2 \frac{m^3_{M}}{(m_q + m_a)^2},
\\&&
\langle \bar{M} ^0 | (\bar{a} P_R  q)(\bar{a} P_L  q) | M^0 \rangle 
 \rangle 
= \frac{1}{24} f_{M}^2 m_{M} - \frac{1}{4} f_{M}^2 \frac{m^3_{M}}{(m_q + m_a)^2}.
\end{eqnarray}

\section{Examples of amplitudes}
\label{sec:examples}

As we explain in Sec.~\ref{sec:eh}, we have classified the amplitudes according to the internal boson lines. First of all we recall that the type of the bosons in the internal lines fix the type of the quark in the internal lines. To write all amplitudes explicitly is not appropriate for the sake of space, hence we will show some of them in order to make our calculations clearer.
 
\subsection{Charged scalar $Y_2$ and charged Goldstone boson $G_V$}
\label{subsec:21}

Here we present the amplitudes arisen when the singly charged scalar $Y^+_2$ and the would-be Goldstone boson related to the vector $V^+_\mu$, denoted by $G^+_V$, are in a box diagram. In this case the fermion in the internal lines are those with charge $-4/3$, i.e., $j_{1,2}$.
We have assumed equal masses for the exotic quarks $j_1, j_2$, i.e. $m_{j_1} = m_{j_2} = m$. All contributions are summarized as follows:
\begin{eqnarray}
i\mathcal{M}_{G_V Y_2} &=& \sum _{l=1} ^2 \sum _{i=1} ^2  \biggl\{ I_V^{\mu \nu}(x_{Y_2}, y_{j_1}) 
\Bigl[(K_1)_{jl}(K_2)^*_{kl}(K_2)_{ji}(K_1)^*_{ki} 
[\gamma_{\mu} P_L]_{jk}[\gamma_{\nu} P_L]_{jk}  + \nonumber \\
& & + (K_1)_{jl}(K_2)^*_{kl}(\tilde{K_2})^*_{ij}(\tilde{K_1})_{ik}  
[\gamma_{\mu} P_L]_{jk}[\gamma_{\nu} P_R]_{jk}
+ (\tilde{K_1})^*_{lj}(\tilde{K_2})_{lk}(K_2)_{ji}(K_1)^*_{ki} 
[\gamma_{\mu} P_R]_{jk}[\gamma_{\nu} P_L]_{jk} + \nonumber \\
& & + (\tilde{K_1})^*_{lj}(\tilde{K_2})_{lk}(\tilde{K_2})^*_{ij}(\tilde{K_1})_{ik} 
[\gamma_{\mu} P_R]_{jk}[\gamma_{\nu} P_R]_{jk} \Bigr]+ \nonumber \\
& & + m^2 I_V (x_{Y_2}, y_{j_1})  \Bigl[
(K_1)_{jl}(\tilde{K_2})_{lk}(K_2)_{ji}(\tilde{K_1})_{ik}
[P_R]_{jk}[P_R]_{jk}+ \nonumber \\
& & + (K_1)_{jl}(\tilde{K_2})_{lk}(\tilde{K_2})^*_{ij}(K_1)^*_{ki}
[P_R]_{jk}[P_L]_{jk}+ 
(\tilde{K_1})^*_{lj}(K_2)^*_{kl}(K_2)_{ji}(\tilde{K_1})_{ik}
[P_L]_{jk}[P_R]_{jk} +\nonumber \\
& & + (\tilde{K_1})^*_{lj}(K_2)^*_{kl}(\tilde{K_2})^*_{ij}(K_1)^*_{ki} 
[P_L]_{jk}[P_L]_{jk} \Bigr] \biggr\}.
\nonumber
\end{eqnarray}
The matrices $K_1,\tilde{K}_1,K_2,\tilde{K}_2$ are given in Eq.~(\ref{a1e3}). 

\subsection{Charged scalar $Y^-_1$ and $W^-$ boson}
\label{subsec:22}

Next, we will show the amplitude when a singly charged scalar $Y^-_1$ and a $W^-$ is exchanged in the box.
In this case the fermions in the internal lines are the SM $u$-type ones. Besides the CKM matrix, the matrices $K_6$ and $\tilde{K}_6$, given in Eq.~(\ref{a1e4}), appear: 
\begin{eqnarray}
i\mathcal{M}_{W Y_1} &=&  g^2 \sum_{i=1}^3
\sum_{l=1}^3
 \Bigl\{ 2 I^{\mu\nu}_{Y_1}(x_i, x_l, x_W) \, (V_{CKM})_{ji}  (V_{CKM})^*_{kl} 
[\gamma _{\mu} P _L]_{jk} \cdot \nonumber \\ 
& & \{ (K_6)_{jl}(K_6)^*_{ki} [\gamma _{\nu} P_L]_{jk}   + 
( \tilde{K_6})_{ik}( \tilde{K_6})^*_{lj}[\gamma_\nu P_R]_{jk}   \} \nonumber 
\Bigr\}
\end{eqnarray}
As we said in Appendix~\ref{sec:integrals}, there are no terms proportional to  $I_{Y_1}(x_i, x_l, x_W)$ once they would be suppressed by a $m_im_l/m^4_{331}$ factor, where $m_i, m_l$ are the masses of the U-quarks.

\subsection{Two charged gauge bosons $V^-$}
\label{subsec:23}

Finally, the most common expression, appearing for instance with W bosons in the framework of SM:
\begin{eqnarray}
i\mathcal{M}_{V} &=& \frac{g^4}{2}  \sum _{i=1}^2 \sum _{l=1}^2 \,I^{\mu \nu}_V(y_{j_1}) (V_L^D)_{jl}(V_L^D)^*_{kl}(V_L^D)_{ji}(V_L^D)^*_{ki} 
\,    [\gamma_{\mu} P_L]_{jk}[\gamma_{\nu} P_L]_{jk},  
\end{eqnarray}
where $V^D_L$ is given in Eq.~(\ref{a1e7}).

\subsection{Penguin diagrams}
\label{subsec:24}

\begin{center}
   	\begin{figure}[h]
   		\subfloat[]{\includegraphics[width=5cm]{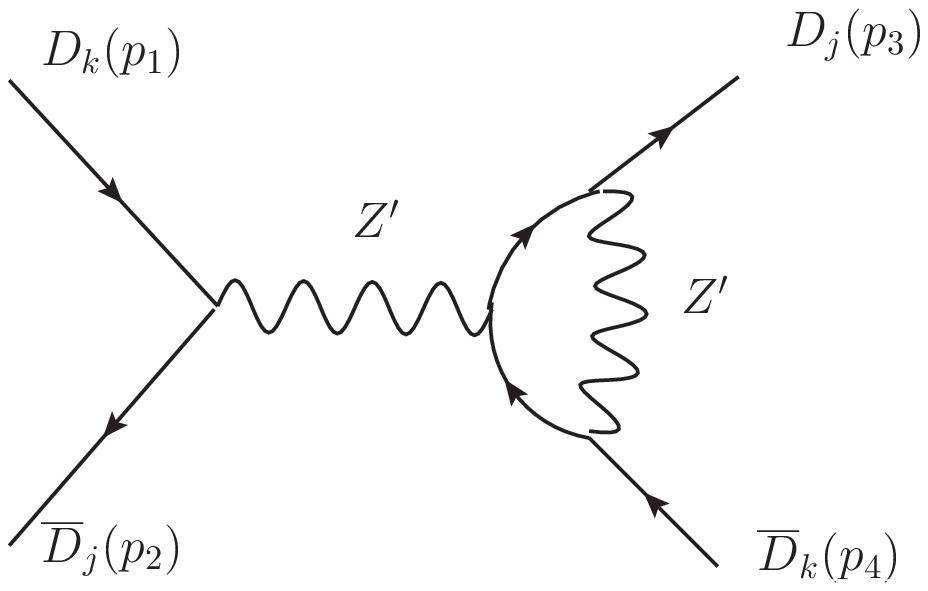}} \qquad
   		\subfloat[]{\includegraphics[width=5cm]{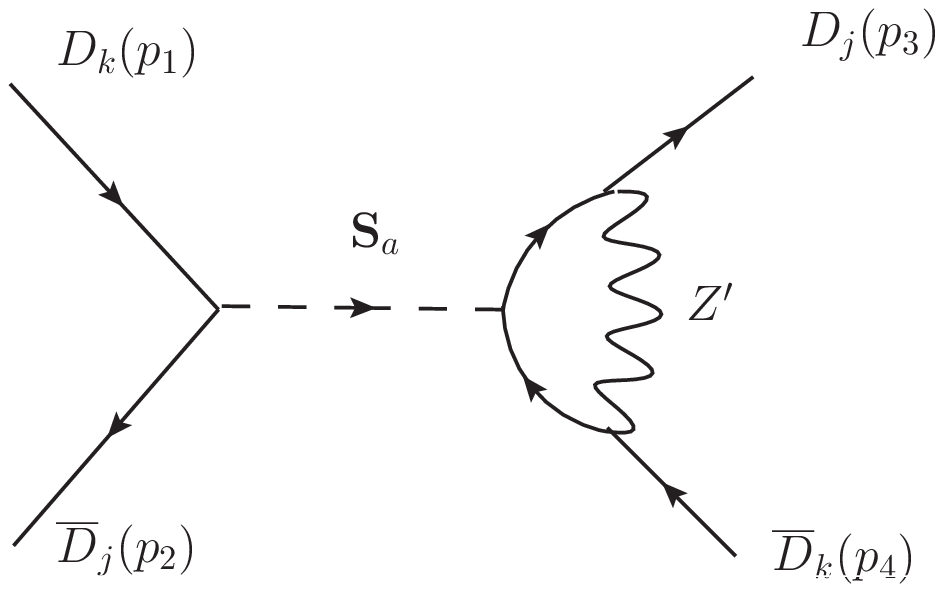}} \\
   		\subfloat[]{\includegraphics[width=5cm]{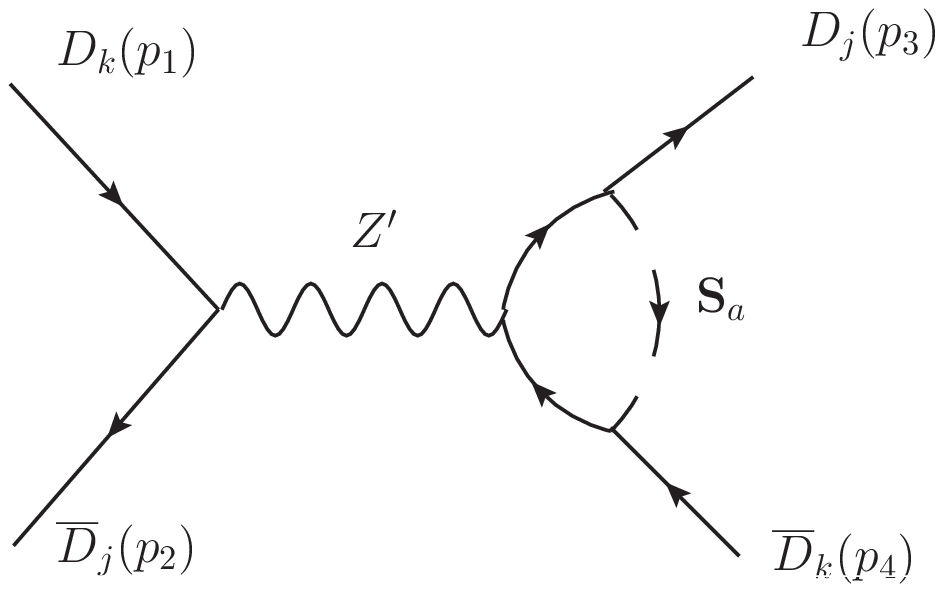}} \qquad
   		\subfloat[]{\includegraphics[width=5cm]{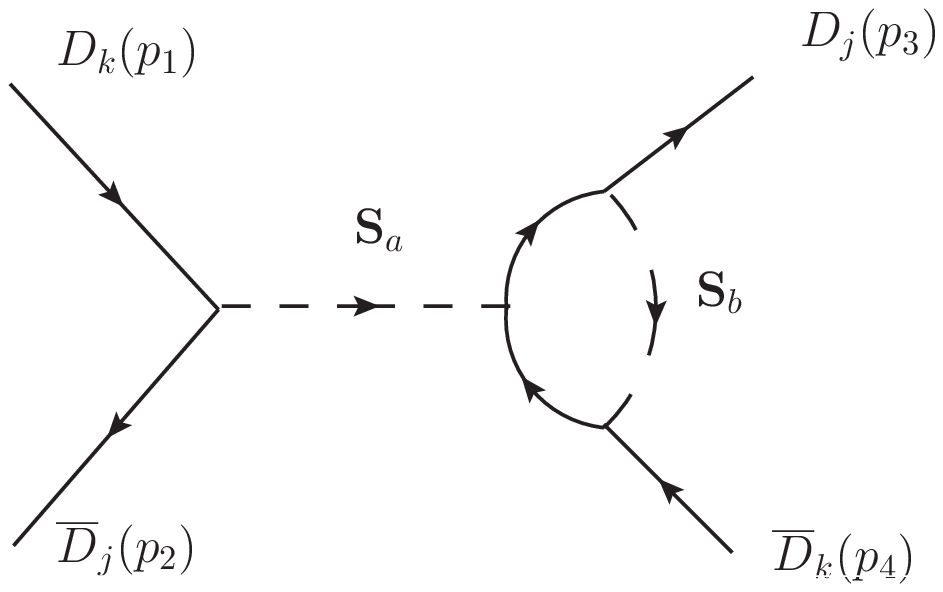}}
   		\caption{The penguin diagrams. These contributions are also negligible by $(\Delta M_M)_{penguin} <10^{-18}$ GeV.}\label{fig16}
   	\end{figure}
   \end{center}

We can still estimate how relevant the penguin diagrams shown in Fig.~\ref{fig16} are. From the matrices $\mathcal{K}^D_h$ and $\mathcal{K}^D_A$, the diagrams with neutral scalars will be negligible, just as their box diagrams were. At first, the same could not be said in the case of the diagram in Fig.~\ref{fig16}(a) which involve two $Z'$, once the $(N^D_{Z'})_{ij}$'s are not small. To estimate it we will neglect the factors proportional to $\frac{m_i m_l}{m_{Z^\prime}^4}$ ($i,l$ are D-quark indices), just as we have done throughout the paper. The remaining contributions can be written as
\begin{equation}
i \mathcal{M}_{10 (a)} = -\sum _{i,l} \frac{2 \beta _{il} g^4}{(4\pi)^2m^2 _{Z'} } 
 [\gamma_{\mu} P_L]_{jk}[\gamma^{\mu} P_L]_{jk}
\int _0 ^1 \int _0 ^{1-x} dxdy   \log \Delta  ,
\label{penguin}
\end{equation}
here $\Delta = m_i^2 x + m^2_l y + m^2 _{Z'}(1-x-y) > 0$ and $\beta_{il} = (K^D_L)_{jk}(K^D_L)_{ji}(K^D_L)_{il}(K^D_L)_{lk}$. In (\ref{penguin}) we have omitted the terms that should be subtracted in the $\overline{MS}$ renormalization scheme. Finally, for a $M_{Z^\prime}$ of 2.4 TeV,  we obtain 
$(\Delta M_M)_p < 10^{-18}$ GeV for all $K,B_{d,s}$ mesons.

\newpage 

   \begin{center}
   	\begin{figure}[h]
   		\subfloat[]{\includegraphics[width=5cm]{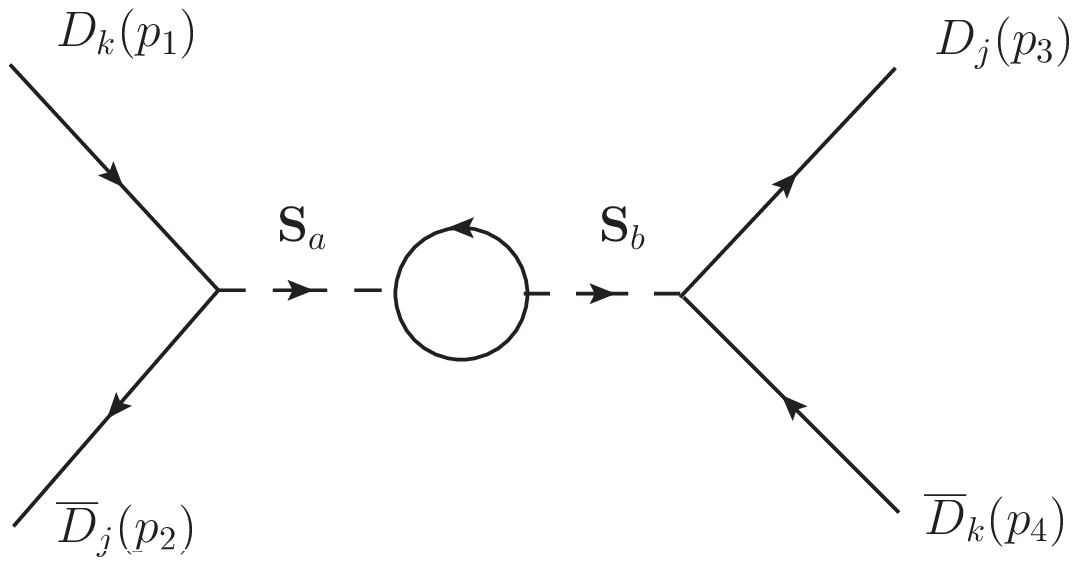}} \qquad
   		\subfloat[]{\includegraphics[width=5cm]{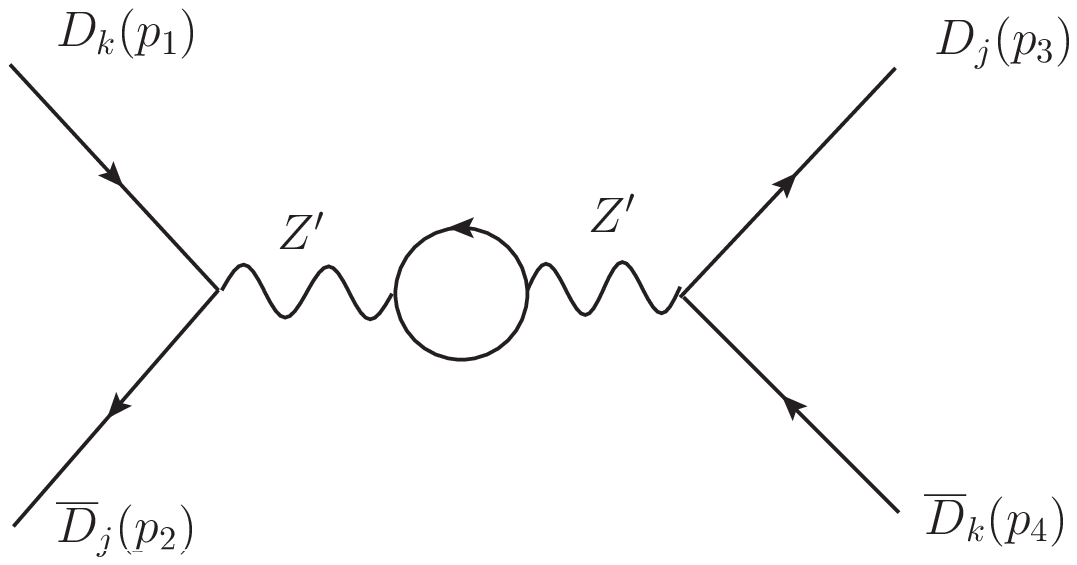}}
   		\caption{The loop corrections to the propagator of neutral scalars and $Z'$.}\label{fig17}
   	\end{figure}
   \end{center} 
  
  \begin{figure}[ht]
  \includegraphics[scale=0.8]{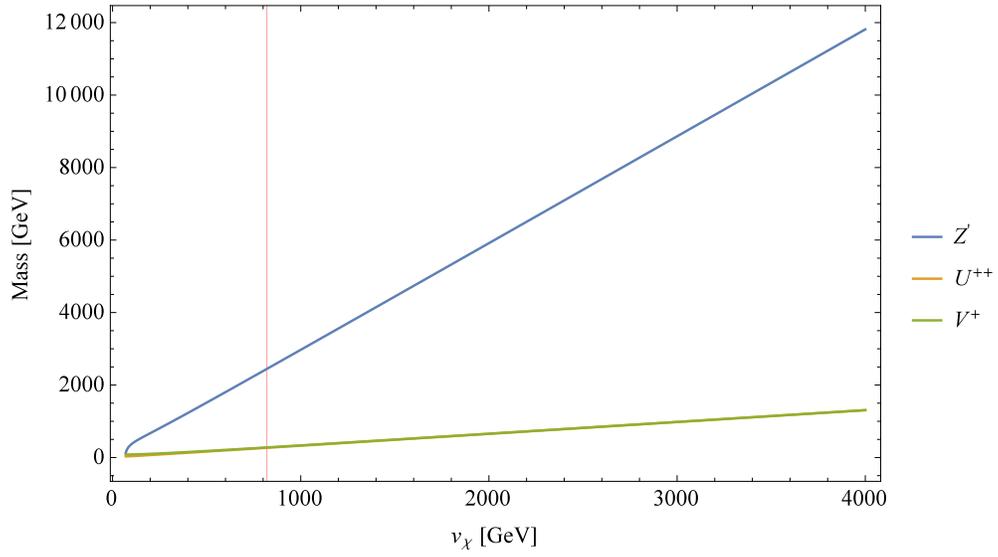}
  \caption{The gauge boson masses as a function of $v_\chi$. The vertical red line is a constraint from $K^0-\bar{K}^0$.} \label{fig10}
  \end{figure}
  
   \begin{figure}[ht]
    \includegraphics[scale=0.8]{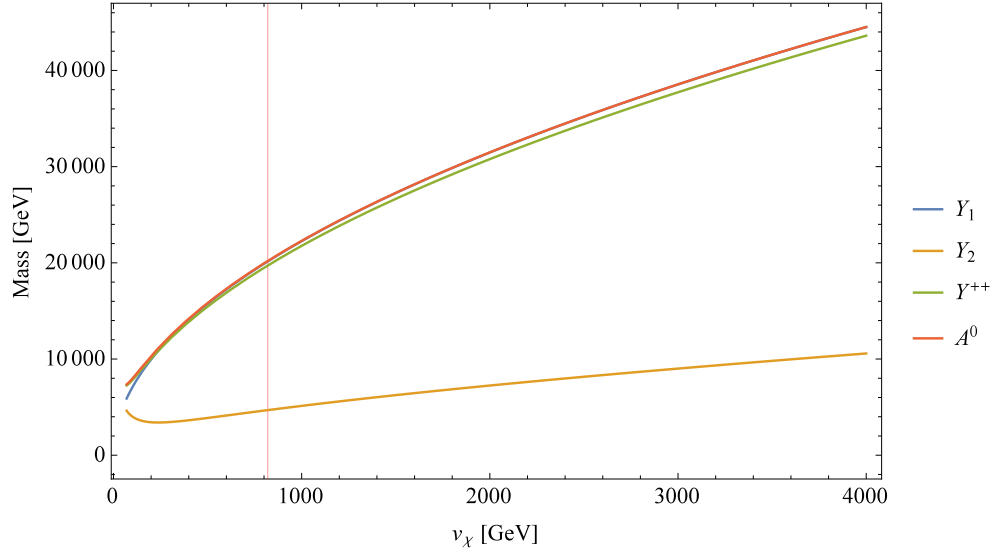}
    \caption{The scalar masses as a function of $v_\chi$. The vertical red line is a constraint from $K^0-\bar{K}^0$.} \label{fig11}
    \end{figure}
  
  \begin{figure}[ht]
  \includegraphics[scale=0.8]{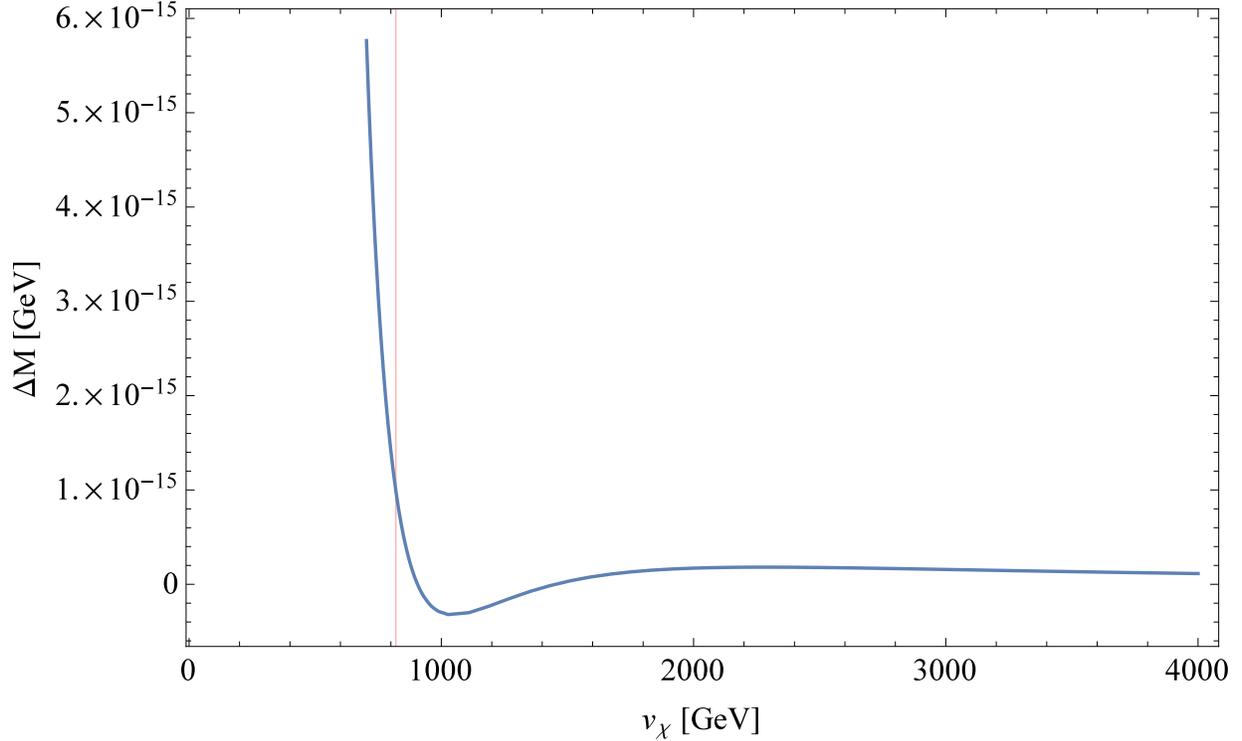}
  \caption{The contributions from box diagrams in the 331HL model to the mass difference of the $K^0$ as a function of $v_\chi$. The vertical red line at $v_\chi = 820$ GeV corresponds to a lower limit in order to obtain the experimental limit $10^{-15}$ GeV to $\Delta m$ in $K^0-\bar{K}^0$.} \label{fig12}
  \end{figure}
 
  \begin{figure}[ht]
   \includegraphics[scale=0.8]{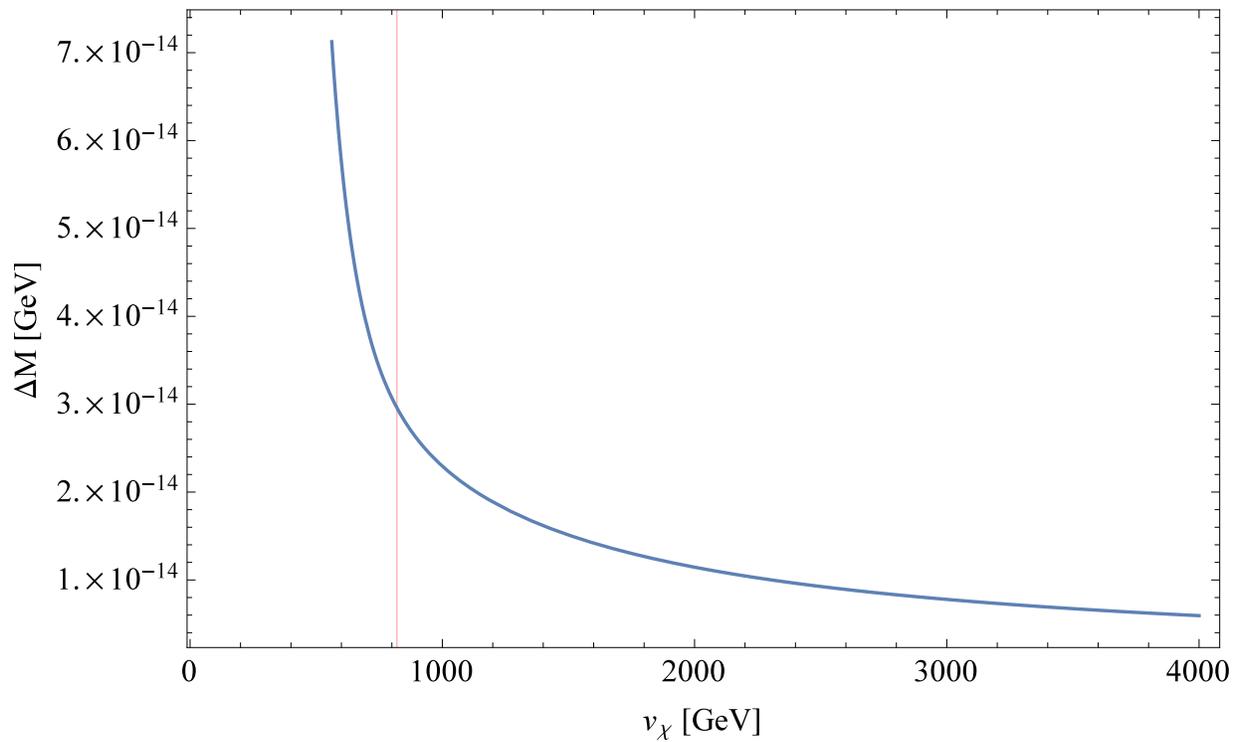}
   \caption{The contributions to the mass difference of the $B^0_d$ as a function of $v_\chi$. The vertical red line corresponds to a lower limit in order to obtain the experimental limit experimental limit $10^{-15}$ GeV to $\Delta m$ in $K^0-\bar{K}^0$.} \label{fig13}
   \end{figure}
   
   \begin{figure}[ht]
      \includegraphics[scale=0.8]{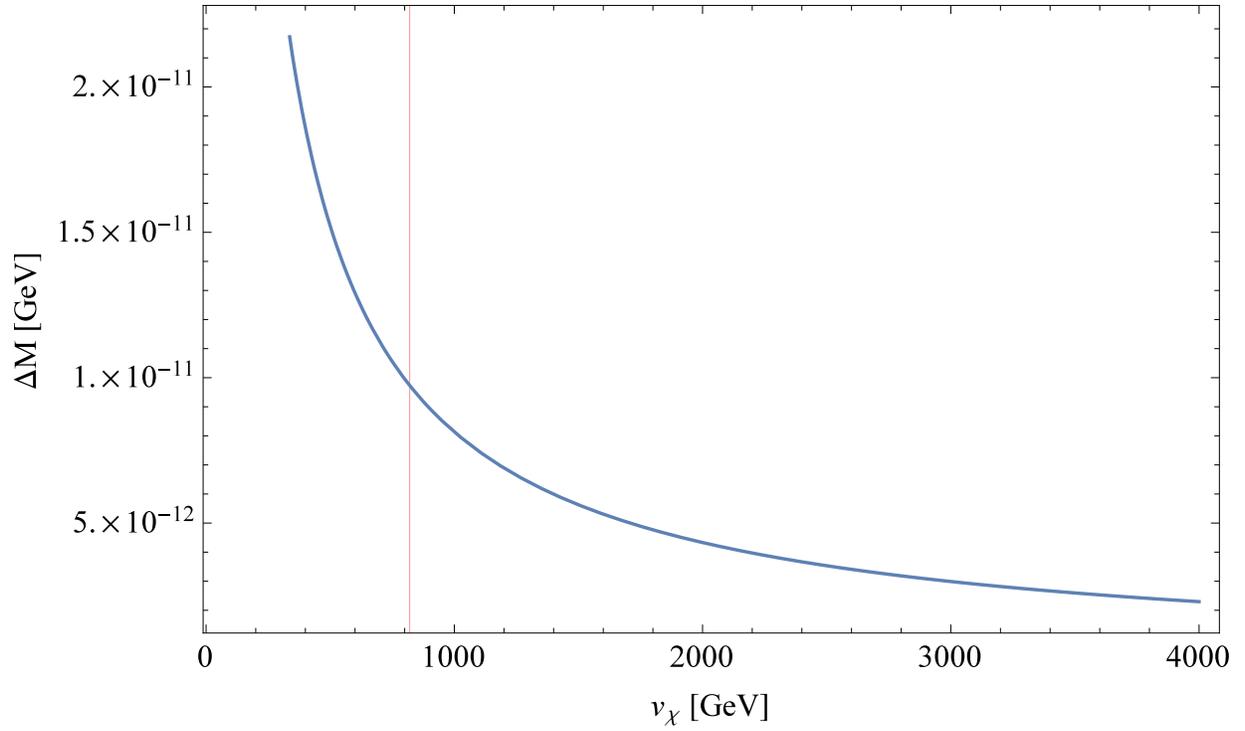}
      \caption{The contributions to the mass difference in $B_s-\bar{B}_s$ as a function of $v_\chi$. The vertical red line corresponds to the experimental limit $10^{-15}$ GeV from $\Delta m$ in $K^0-\bar{K}^0$.} \label{fig14}
      \end{figure}

\end{document}